\RequirePackage{fix-cm}
\documentclass[twocolumn,epjc3]{svjour3}

\usepackage{atlasphysics} 
                           
\usepackage{bm}
\usepackage{multirow}
\usepackage{array}
\usepackage{cite}
\usepackage{subfigure}

\usepackage[hyperindex,breaklinks]{hyperref}

\usepackage{preprintcover} 

\PreprintCoverPaperTitle{Measurement of the flavour composition of dijet events in $\bm{pp}$ collisions at $\bm{\rts=7\TeV}$ with the ATLAS detector}  

\PreprintIdNumber{CERN-PH-EP-2012-255} 

\PreprintCoverAbstract{This paper describes a measurement of the flavour composition of dijet events produced in $pp$ collisions 
at $\rts=7\TeV$ using the ATLAS detector. 
The measurement uses the full 2010 data sample, corresponding to an integrated luminosity of $39$~$\,\ipb$.
Six possible combinations of light, charm and bottom jets are identified in the dijet events,
where the jet flavour is defined by the presence of bottom, charm or solely light flavour hadrons in the jet. 
Kinematic variables, based on the properties of displaced decay vertices and optimised for jet flavour identification, 
are used in a multidimensional template fit to measure the fractions of these dijet flavour states 
as functions of the leading jet transverse momentum in the range $40\GeV$ to $500\GeV$ and jet rapidity $|y| < 2.1$. 
The fit results agree with the predictions of leading- and next-to-leading-order calculations, 
with the exception of the dijet fraction composed of bottom and light flavour jets,  
which is underestimated by all models at large transverse jet momenta.
The ability to identify jets containing two $b$-hadrons, originating from e.g. gluon splitting, is demonstrated.
The difference between bottom jet production rates in leading and subleading jets 
is consistent with the next-to-leading-order predictions.} 

\PreprintJournalName{European Physical Journal C}

\smartqed  
\RequirePackage{graphicx}
\RequirePackage{mathptmx}      
\journalname{Eur. Phys. J. C}

\begin{document}

\title{Measurement of the flavour composition of dijet events in $\bm{pp}$ collisions at $\bm{\rts=7\TeV}$ with the ATLAS detector
}

\author{The ATLAS Collaboration}

\institute{}

\date{Received: date / Accepted: date}

\maketitle

\begin{abstract}
This paper describes a measurement of the flavour composition of dijet events produced in $pp$ collisions 
at $\rts=7\TeV$ using the ATLAS detector. 
The measurement uses the full 2010 data sample, corresponding to an integrated luminosity of $39$~$\,\ipb$.
Six possible combinations of light, charm and bottom jets are identified in the dijet events,
where the jet flavour is defined by the presence of bottom, charm or solely light flavour hadrons in the jet. 
Kinematic variables, based on the properties of displaced decay vertices and optimised for jet flavour identification, 
are used in a multidimensional template fit to measure the fractions of these dijet flavour states 
as functions of the leading jet transverse momentum in the range $40\GeV$ to $500\GeV$ and jet rapidity $|y| < 2.1$. 
The fit results agree with the predictions of leading- and next-to-leading-order calculations, 
with the exception of the dijet fraction composed of bottom and light flavour jets,  
which is underestimated by all models at large transverse jet momenta.
The ability to identify jets containing two $b$-hadrons, originating from e.g. gluon splitting, is demonstrated.
The difference between bottom jet production rates in leading and subleading jets 
is consistent with the next-to-leading-order predictions.
\end{abstract}

\section{Introduction}
\label{sec:Introduction}
\sloppy

A study of the production of jets containing bottom and charm hadrons, which are likely to have originated from bottom or charm quarks, 
is of strong interest for an understanding of Quantum Chromodynamics (QCD).
Charm and bottom quarks have masses significantly above the QCD scale, $\rm{\Lambda_{QCD}}$, 
and hence low energy hadronization effects should not influence the total cross section and the distributions of the charm and bottom hadrons.
In this approximation, properties of the jets containing heavy flavour hadrons are expected to be described accurately using perturbative calculations. 
A measurement of the production features of these jets can thus shed light on the details of the underlying QCD dynamics.

Several mechanisms contribute to heavy flavour quark production, such as quark-antiquark pair creation in the hard interaction or in the parton showering process.
While the former is calculable in a perturbative approach, 
the latter may require additional non-perturbative corrections or different approaches such as a heavy quark mass expansion.
In inclusive heavy flavour jet cross-sections, the contribution from gluon splitting in the final state parton showering could be identified 
by looking for two heavy flavour hadrons in a jet, 
but the different mechanisms for prompt heavy flavour quark production in the hard interaction remain indistinguishable. 
This complicates a comparison with theoretical calculations.  
A more exclusive study of the production of dijet events containing heavy flavour jets allows 
the different prompt heavy flavour quark creation processes to be separated, in addition to the gluon splitting contribution.
For example, the dominant QCD production mechanisms are different for pairs of bottom flavour jets and pairs consisting of one bottom and one light jet.
In this context, a measurement of the flavour composition of dijet events  
provides more detailed information about the different QCD processes involving heavy quarks. 

The dijet system can be decomposed into six flavour states based on the contributing jet flavours. 
The jet flavour is defined by the flavour of the heaviest hadron in the jet.  
A light jet originates from fragmentation of a light flavour quark ($u$, $d$ and $s$) or gluon and does not contain any bottom or charm hadrons.
Three of these dijet states are the symmetric bottom+bottom ($b\bar b$), charm+charm ($c\bar c$) and light+light jet pairs.
The three other combinations are the flavour-asymmetric bottom+light, charm+light and bottom+charm jet pairs.
In the following discussion, these six dijet flavour states will be denoted $BB$, $CC$, $UU$, $BU$, $CU$, $BC$, where $U$ stands for light,
$C$ for charm and $B$ for bottom jet. 

Inclusive bottom jet and $b \bar b$ production in hadronic collisions have been studied by several experiments~\cite{BBarD0,BInclD0,BBbarCDF,BBarCDFConf,BBarSPS} in the past,
see also a review~\cite{BBarReview} and references therein.
Recently CMS published cross-sections for inclusive bottom jet production~\cite{BInclCMS}, 
$b\bar b$ decaying to muons~\cite{BBarCMS} and bottom hadron production~\cite{BHadCMS}, as well as $B \bar B$ angular correlations~\cite{BBAngCMS}.
The $b \bar b$ cross-section was also measured by LHCb~\cite{BBarLHCb}. 
ATLAS published a measurement of the $b \bar b$ cross-section in proton-proton collisions at $\sqrt{s}=7\TeV$~\cite{BBarAtlasPaper}, employing explicit 
$b$-jet identification ($b$-tagging). 
However, the $b \bar b$ final state constitutes only a small fraction
of the total heavy flavour quark production in dijet events, and the inclusive bottom cross-section contains a significant contribution from multijet states.
This paper presents a simultaneous measurement of all six dijet
flavour states, including those with charm. The $BC$, $CC$ and $CU$ dijet production at the LHC is studied for the first time.
This approach provides more detailed information about the contributing QCD processes and challenges the theoretical description of the 
underlying dynamics employed in QCD Monte Carlo simulations. 

The analysis procedure exploits reconstructed secondary vertices inside jets.
Since kinematic properties of secondary vertices depend on the jet flavour, a measurement of the individual contributions of each flavour 
can be made by employing a fit using templates of kinematic variables.
No explicit $b$-tagging is used, i.e. no flavours are assigned to individual jets.
The excellent separation of charm and bottom flavoured jets in the ATLAS detector is demonstrated in the analysis.

The analysis uses the data sample collected by ATLAS at $\rts=7\TeV$ in 2010, corresponding to an integrated luminosity of $39$~$\ipb$. 
The prescale settings of the different single-jet triggers used in the analysis varied with luminosity such that 
the actual recorded luminosity is dependent on the transverse momentum $\pT$ of the leading jet.  

This paper is organised as follows. 
The ATLAS detector is briefly described  in Sect.~\ref{sec:Detector}.
Section~\ref{sec:Selection} describes the event and jet selection procedure for data and Monte Carlo simulation.
Section~\ref{sec:Mc} summarises the Monte Carlo simulation. 
Section~\ref{sec:Predictions} discusses the theoretical predictions for the flavour composition of dijet events.
The reconstruction of secondary vertices in jets as well as the kinematic templates for the flavour analysis are presented in Sect.~\ref{sec:SVreco}.
A detailed account of the analysis method is given in Sect.~\ref{sec:Method}. 
In Sect.~\ref{sec:Results} the results of the analysis are presented and systematic uncertainties are discussed.

\fussy

\section{The ATLAS detector}
\label{sec:Detector}
The ATLAS detector~\cite{AtlasDetectorPaper} was designed to allow the study of a wide range of physics processes at LHC energies. 
It consists of an inner tracking detector, surrounded by an electromagnetic calorimeter, hadronic calorimeters and a muon spectrometer.
For the measurements presented in this paper, the tracking devices, the calorimeters and the trigger system are of particular importance. 

\begin{table*}[!hbt]
 \begin{center}
 \begin{tabular}{ccccccc}
 \hline
 Leading jet $\pt$ [\GeV]  & 40--60  & 60--80  & 80--120 & 120--160  & 160--250  & 250--500 \\
 \hline
Subleading jet $\pt$ [\GeV] & 30--60 & 40--80 & 50--120 &  75--160 & 100--250 & 140--500 \\
 \hline
 Number of events &  304103 & 251406 &  887185 &  660168 & 242979 & 146117 \\
 \hline
$\int L\mathrm{d}t\,$[$\inb$]  &   70    & 247  & 1880  & 8640  & 8640   &  38700 \\
\hline
\end{tabular}
\end{center}
\caption{Kinematic boundaries, together with the numbers of selected dijet events 
         and the corresponding integrated luminosities for each leading jet $\pT$ bin.}
\label{tables:PtBinsAndEvents}
\end{table*}

The innermost detector, the tracker, is divided into three parts: the silicon pixel detector, the closest layer lying \linebreak $5.05\,$cm from the beam axis, 
the silicon microstrip detector and the transition radiation
tracker, with the outermost layer situated at $1.07\,$m from the beam axis. 
These offer full coverage in the azimuthal angle $\phi$ and a coverage in pseudorapidity of $|\eta|<2.5$~\footnote{ATLAS uses a right-handed coordinate system 
with its origin at the nominal interaction point (IP) in the centre of the detector and the $z$-axis along the beam pipe. 
The $x$-axis points from the IP to the centre of the LHC ring, and the $y$-axis points upward. 
Cylindrical coordinates $(r,\phi)$ are used in the transverse plane, $\phi$ being the azimuthal angle around the beam pipe. 
The pseudorapidity is defined in terms of the polar angle $\theta$ as $\eta=-\ln\tan(\theta/2)$.}. 
The tracker is surrounded by a solenoidal magnet of $2\,$T, which bends the trajectories of charged particles 
so that their transverse momenta can be measured.
The liquid argon and lead electromagnetic calorimeter covers a pseudorapidity range of $|\eta| < 3.2$.
It is surrounded by the hadronic calorimeters, made of scintillator tiles 
and iron in the central region ($|\eta| < 1.7$) and of copper/tungsten and liquid argon in the endcaps ($1.5 < |\eta| < 3.2$). 
A forward calorimeter extends the coverage to $|\eta| < 4.9$.
The muon spectrometer comprises three layers of muon chambers for track measurements and triggering. It uses a toroidal magnetic field 
with a bending power of $1$--$7.5\,$Tm and provides precise tracking information
in a range of $|\eta| < 2.7$.

The ATLAS trigger system~\cite{AtlasDetectorPaper} uses three consecutive levels: level 1 (L1), level 2 (L2) and event filter (EF). 
The L1 triggers are hardware-based and use coarse detector information to identify regions of interest,
whereas the L2 triggers are based on fast online data reconstruction algorithms. 
Finally, the EF triggers use offline data reconstruction algorithms. 
This study uses single-jet triggers.

\section{Event and jet selection}
\label{sec:Selection}

\sloppy

Selected events are required to have at least one reconstructed primary vertex candidate.
A candidate vertex must have at least 10 tracks with transverse momentum $\pt>150\MeV$ associated to it, to ensure the quality of the vertex fit.
If several vertex candidates are reconstructed, the one with the largest sum of the squared transverse momenta of associated tracks
is considered to be the main interaction vertex and used as the primary vertex in the following. 

Jets are reconstructed using the anti-$k_t$ algorithm with \\ a jet radius parameter $R=0.4$~\cite{anti_kt}. 
Topological clusters of energy deposits in the calorimeters
are used as input for the clustering algorithm.
Tracks within a cone of $\Delta R = \sqrt{(\Delta \varphi) ^{2} + (\Delta \eta) ^{2}}=0.4$ around the jet axis are assigned to the jet.
Only jets with a transverse momentum of $\pt>30\GeV$ and a rapidity of $|y|<2.1$ are considered.
Jets in this rapidity range are fully contained in the tracker acceptance region,
such that track and vertex reconstruction inside jets are not affected by the boundaries of the tracker acceptance.
Jets are furthermore required to pass a quality selection~\cite{ATLAS-CONF-2010-038,JetEnergyAtlas}  
that removes jets mimicked by noisy calorimeter cells
or those that stem from non-collision backgrounds. 
Finally, the two jets with highest $\pT$ in the analysis acceptance
are required to have an angular separation in azimuth of $\Delta\varphi>2.1\,$rad, 
i.e. to be consistent with a back-to-back topology. This cut removes events in which one of the leading jets is produced by final-state hard gluon emission or jet splitting in the reconstruction.

 The full data sample is split into six bins in the transverse momentum $\pT$ of the leading jet.
The bin boundaries correspond to the $99\,$\% efficiency thresholds of the various single-jet triggers~\cite{atlas_performance}. 
For events passing the trigger requirement, the leading and subleading jets have to fulfil pairwise-specific $\pt$ conditions
that are summarised in \mbox{Table~\ref{tables:PtBinsAndEvents}}. The numbers of events selected in each leading jet $\pt$ bin are shown in Table~\ref{tables:PtBinsAndEvents}, together with the corresponding integrated luminosities.

\fussy

\section{Monte Carlo simulation}
\label{sec:Mc}
\sloppy
Dijet events are simulated using  
{\sc Pythia}~6.423~\cite{pythia64} for the baseline template construction, parameter estimation
and Monte Carlo (MC) comparisons. This leading-order (LO) generator is
based on parton matrix-element calculations for $2\rightarrow2$ processes 
and a string hadronisation model. 
Modified leading-order MRST LO*~\cite{Sherstnev:2007nd} parton distribution functions are used in the simulation.
Samples of dijet events were generated using a specific set of generator parameters, known as the ATLAS Minimum Bias Tune 1 (AMBT1)~\cite{ATLAS-CONF-2010-031}.

For the study of systematic effects
and for the interpretation of the final results, other Monte Carlo
samples are utilised. The main cross-check study is performed using the Herwig++~2.4.2~\cite{herwigpp} generator.
The other LO samples used are {\sc Pythia} with the next-to-leading-order (NLO) CTEQ~6.6~\cite{Nadolsky:2008zw} parton distribution functions and
Herwig~6.5~\cite{herwig} used with {\sc Jimmy}~4~\cite{jimmy1,jimmy2} for the simulation of multiple parton interactions, using a specific ATLAS Underlying Event Tune (AUET1)~\cite{ATLAS:1303025}.
The possible influence of multiple proton-proton interactions within the same bunch crossing is studied by adding 
minimum bias events, customised to the beam conditions of the
2010 LHC run at $7\TeV$, to each {\sc Pythia} event.

The {\sc Pythia}~6.423+{\sc EvtGen}~\cite{EvtGen} event generator, 
using charm and bottom decay matrix elements with all sequential decay correlations and measured branching ratios, where available,
is utilised for the simulation of the physics of bottom and charm hadron decays.
It will be called {\sc Pythia}+{\sc EvtGen} in the rest of the paper.

The NLO generator {\sc Powheg}~\cite{pow,powbis,powter,powdij} is used
to interpret the analysis results.
In {\sc Powheg}, the parton distribution function set used for the event generation is MSTW 2008 NLO~\cite{mstw2008} and the parton shower generator is {\sc Pythia}.

In order to compare Monte Carlo predictions with data, ``truth-particle'' jets are used. 
They are defined by the anti-$k_t$ $R=0.4$ algorithm 
using only stable particles with a lifetime longer than $10\,$ps in the Monte Carlo event record.
Muons and neutrinos do not contribute significantly to the jet energy in data. Therefore, they are also excluded from the truth-particle jets,
to avoid having to correct for the missing jet energy in data.

The flavour of jets is assigned in the Monte Carlo simulation by labelling a jet as a $b$-jet if a bottom hadron with $\pT>5\GeV$ is found
within a cone 
\mbox{$\Delta R = 0.3$} around the jet axis.
If no bottom hadron is present but a charm hadron is found using the same requirements, then the jet is labelled as a $c$-jet.
All other jets are labelled as light jets. If two bottom hadrons with $\pT>5\GeV$ are found within a cone of size $\Delta R=0.3$
the jet is labelled as a $b$-jet with two bottom hadrons, and
similarly for $c$-jets with two charm hadrons.

The particle four-momenta are passed through the full simulation~\cite{atlas_simulation} of the ATLAS detector, 
which is based on {\sc Geant}4~\cite{geant4}.
The simulated events are reconstructed and selected using the same analysis chain as for data.
After the dijet event selection, the Monte Carlo events are reweighted in each analysis $\pT$ bin to match the observed leading and subleading jet $\pT$ spectra.
Any remaining discrepancies in the rapidity distributions between data and simulation are small and are included as sources of systematic uncertainty, as detailed in Sect.~\ref{sec:ResultsSysUncert}.

\fussy

\begin{figure*}[!hbt]
\begin{center}
\subfigure[]{
\includegraphics[width=0.4\textwidth]{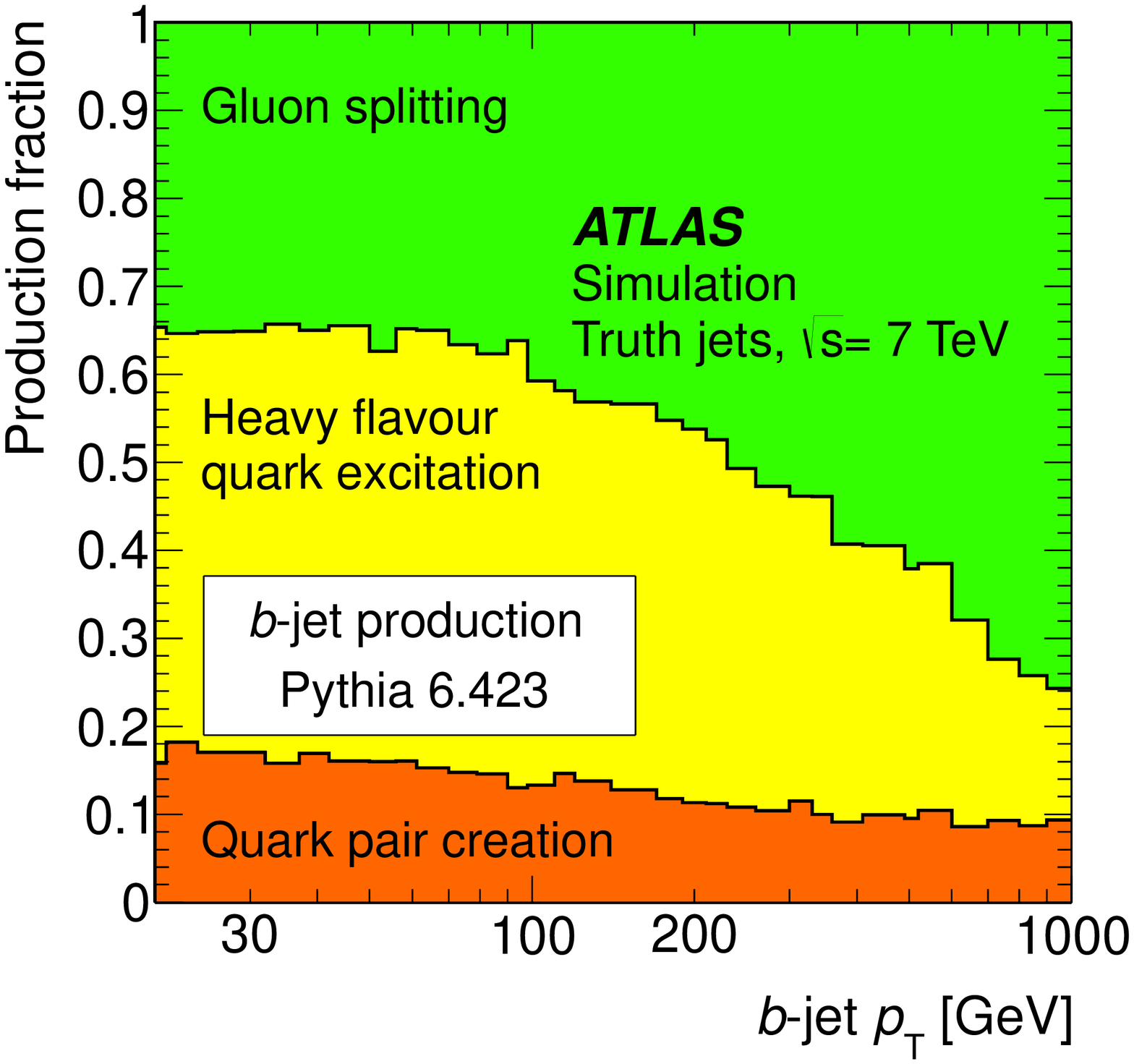}
\label{figures:BProda}
}
\subfigure[]{
\includegraphics[width=0.4\textwidth]{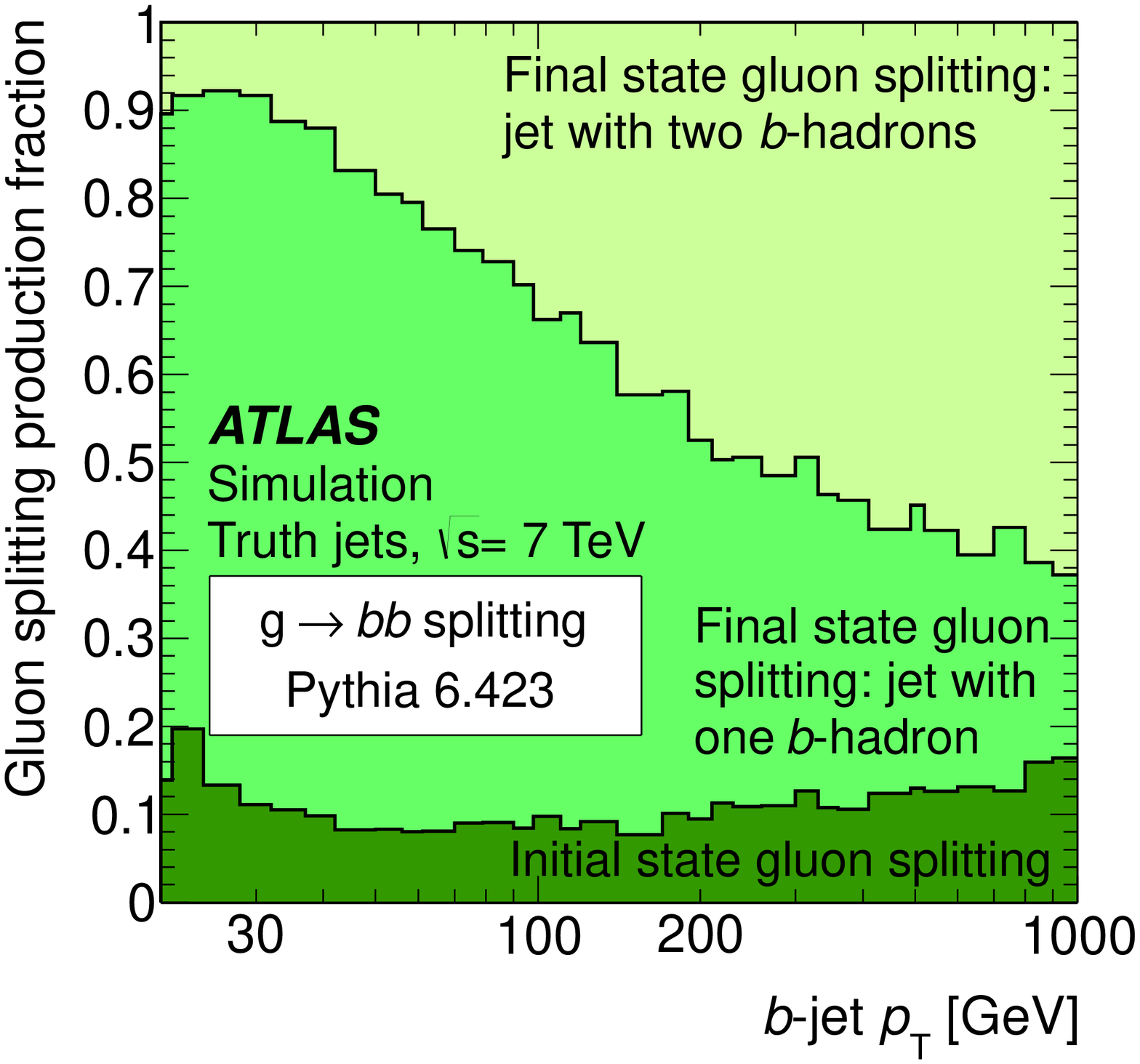}
\label{figures:BProdb}
}
 \caption{The contributions of the different production processes to inclusive $b$-jet production in $7\TeV$ $pp$ collisions 
 are shown as a function of $b$-jet $\pT$, as given by {\sc Pythia}~6.423 and obtained for truth-particle jets. The plot on the left (a) shows the contribution of quark pair creation,
 heavy flavour quark excitation and gluon splitting; the plot on the right (b) shows the different processes contributing to gluon splitting, namely initial- and final-state gluon splitting, 
 the latter leading to jets with one or two $b$-hadrons. Truth-particle jets are reconstructed with the anti-$k_t$ $R=0.4$ algorithm in the $|y|<2.1$ rapidity region. } 
\label{figures:BProd}
\end{center}
\end{figure*}

\section{Theoretical predictions}
\label{sec:Predictions}
\subsection{Heavy flavour production}
\label{subsec:HFProd}

Following the discussion in \cite{Norrbin:2000zc}, heavy flavour quark production in hadronic collisions may be subdivided into three classes
depending on the number of heavy quarks participating in the hard scattering. Hard scattering is defined as the $2 \rightarrow 2$ subprocess
with the largest virtuality (or shortest distance) in the hadron-hadron interaction. In the following, $Q$ stands for a heavy flavour quark, $q$ for a light flavour quark and $g$ for a gluon:

\begin{itemize}
\item \emph{Quark pair creation}: two heavy quarks are produced in the hard subprocess. 
     At leading order this is described by $gg \rightarrow Q \bar Q$ and $q \bar q \rightarrow Q \bar Q$.
\item \emph{Heavy flavour quark excitation}: a single heavy flavour quark from the sea of one hadron scatters 
    against a parton from another hadron, denoted $g Q \rightarrow g Q$ and $q Q \rightarrow q Q$, respectively.
    Alternatively, the heavy flavour quark excitation process
    can be depicted as an initial-state gluon splitting into a heavy quark pair,
    where one of the heavy quarks subsequently enters the hard subprocess. 
\item \emph{Gluon splitting}: in this case heavy quarks do not parti\-ci\-pate in the hard subprocess at all,
             but are produced in $g \rightarrow Q \bar Q$ branchings in the parton shower.
\end{itemize}

\sloppy
The relative contributions of the different heavy flavour quark production mechanisms
to inclusive $b$-jet production are shown in Fig.~\ref{figures:BProda} for simulated proton-proton collisions at $7\TeV$.
The fractions are calculated for anti-$k_t$ jets in a rapidity range of $|y|<2.1$ with the {\sc Pythia}~6.423~\cite{pythia64} generator. 
Figure~\ref{figures:BProdb} shows the decomposition of the gluon splitting process into initial- and final-state gluon splitting, the latter leading to jets with one or two $b$-hadrons.

The above classification is not strict 
but can be used as a basis for gaining a qualitative understanding of
the features of heavy flavour quark production.
Pair creation of heavy flavour quarks gives an insight into perturbative QCD with massive quarks.
The back-to-back requirement used in the analysis reduces the contribution of NLO QCD effects
to the jet-pair cross-sections with two heavy flavour jets, $BB$ and $CC$.
The heavy flavour quark excitation process, on the other hand, is sensitive to the heavy flavour components of the parton distribution functions of the proton.
It produces mainly flavour asymmetric $BU$ and $CU$ jet pairs.
The gluon splitting mechanism is sensitive to non-perturbative QCD
dynamics and also contributes significantly to the mixed flavour jet pair states, i.e. $BU$ and $CU$.
However, this contribution is different from heavy flavour quark excitation 
because it creates a heavy quark-antiquark pair. The jet reconstruction algorithm either includes 
both heavy quarks in a single jet or misses one of them, thus reducing the reconstructed jet energy and its fraction taken by the 
remaining quark.
The two possibilities result in different kinematic properties of the
reconstructed secondary vertices in these jets, which can be exploited 
for the separation of gluon splitting from the heavy flavour quark excitation contribution.

\fussy

\begin{figure}[!hbt]
\begin{center}
  \includegraphics[width=0.8\columnwidth]{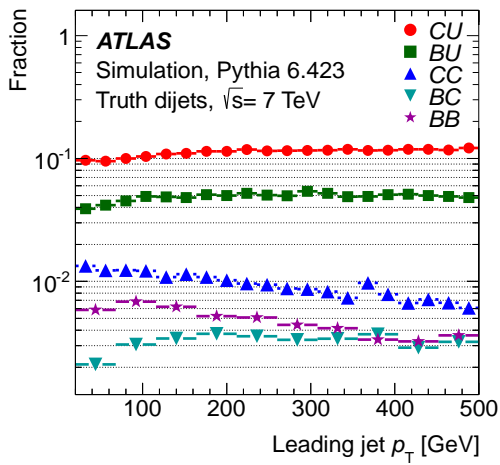}   
  \caption{{\sc Pythia}~6.423 predictions for different bottom and charm dijet fractions as a function of leading jet $\pt$, obtained
          for truth-particle jet pairs, where the jets are back-to-back and have $\pT>20\GeV$ in the $|y|<2.1$ rapidity region.} 
\label{figures:FracPYTH}
\end{center}
\end{figure}

To compare the predictions of theoretical models with data, the truth-particle jets defined in Section~\ref{sec:Mc} are used in the analysis.
The truth-particle dijet system is defined as the two truth-particle
jets with the highest $\pT$ in the $|y|<2.1$ rapidity range, required
to be consistent with a back-to-back topology, $\Delta\varphi>2.1\,$rad, with both the leading and subleading jets having $\pT>20\GeV$. 
 
\sloppy
The leading-order predictions for flavour jet production in truth-particle dijet events are illustrated in Fig.~\ref{figures:FracPYTH},
where the ratio of different heavy+heavy and heavy+light dijet cross-sections to the total dijet cross-section is shown for $|y|<2.1$ 
as a function of leading jet $\pT$, for $7\TeV$ $pp$ collisions as predicted by {\sc Pythia}~6.423. 
Heavy flavour jets in the dijet system are mainly produced in the $BU$ and $CU$ combinations. 
{\sc Pythia}~6.423 predicts a slow decrease of the $BB$ and $CC$ fractions
and an increase of the $BU$ and $CU$ jet fractions as a function of the leading jet $\pT$. 
The mixed $BC$ fraction increases with jet $\pT$ and becomes equal to the $BB$ fraction above $\sim 350\GeV$.

\fussy

\subsection{Differences in heavy flavour rates in leading and subleading jets}
\label{sec:Asymmetry}
\begin{figure}[!hbt]
\begin{center}
\subfigure[]{
\includegraphics[viewport = 12 0 540 728, width=0.2307\textwidth]{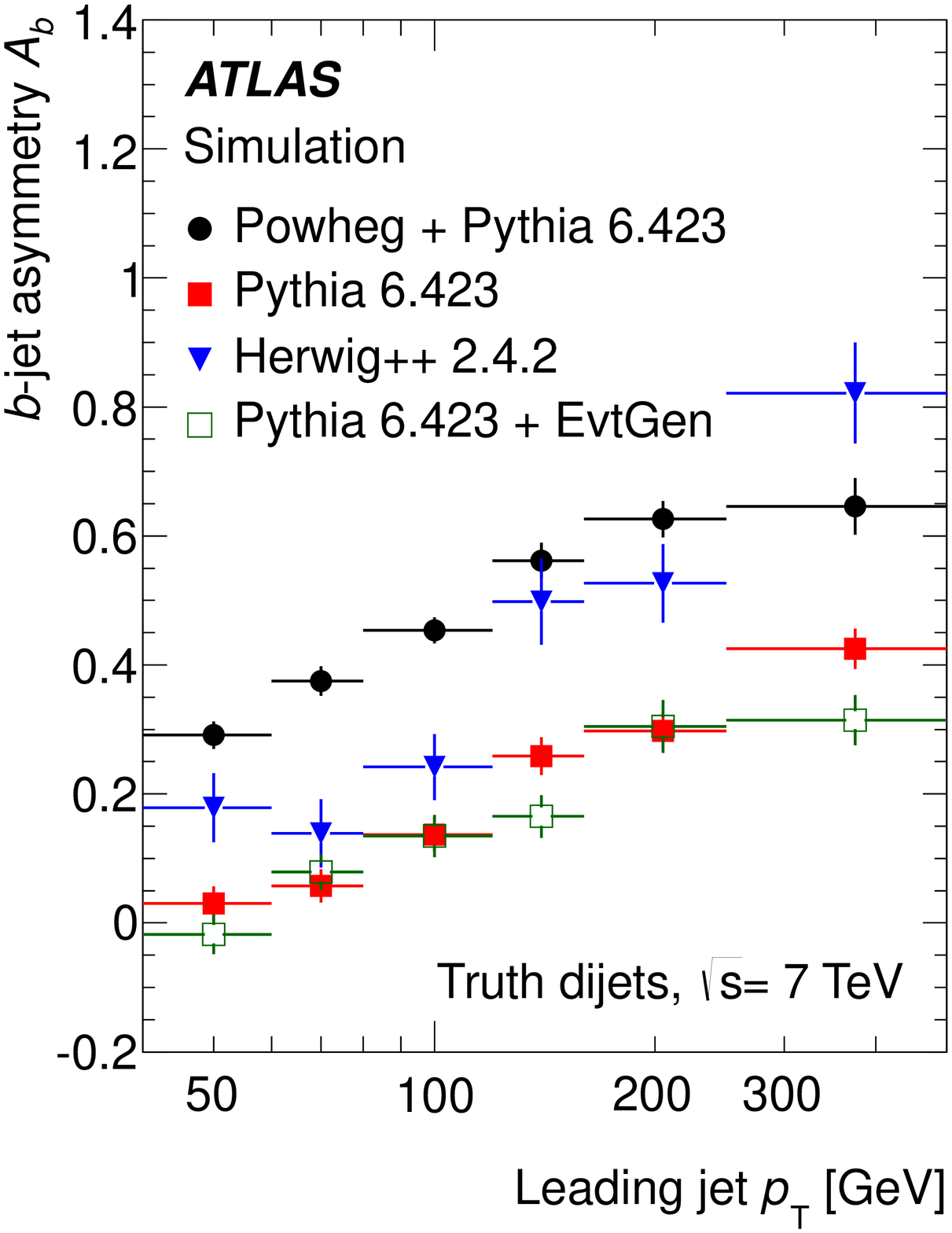}
\label{figures:asymflav1}
}
\subfigure[]{
\includegraphics[viewport = 12 0 540 728, width=0.2307\textwidth]{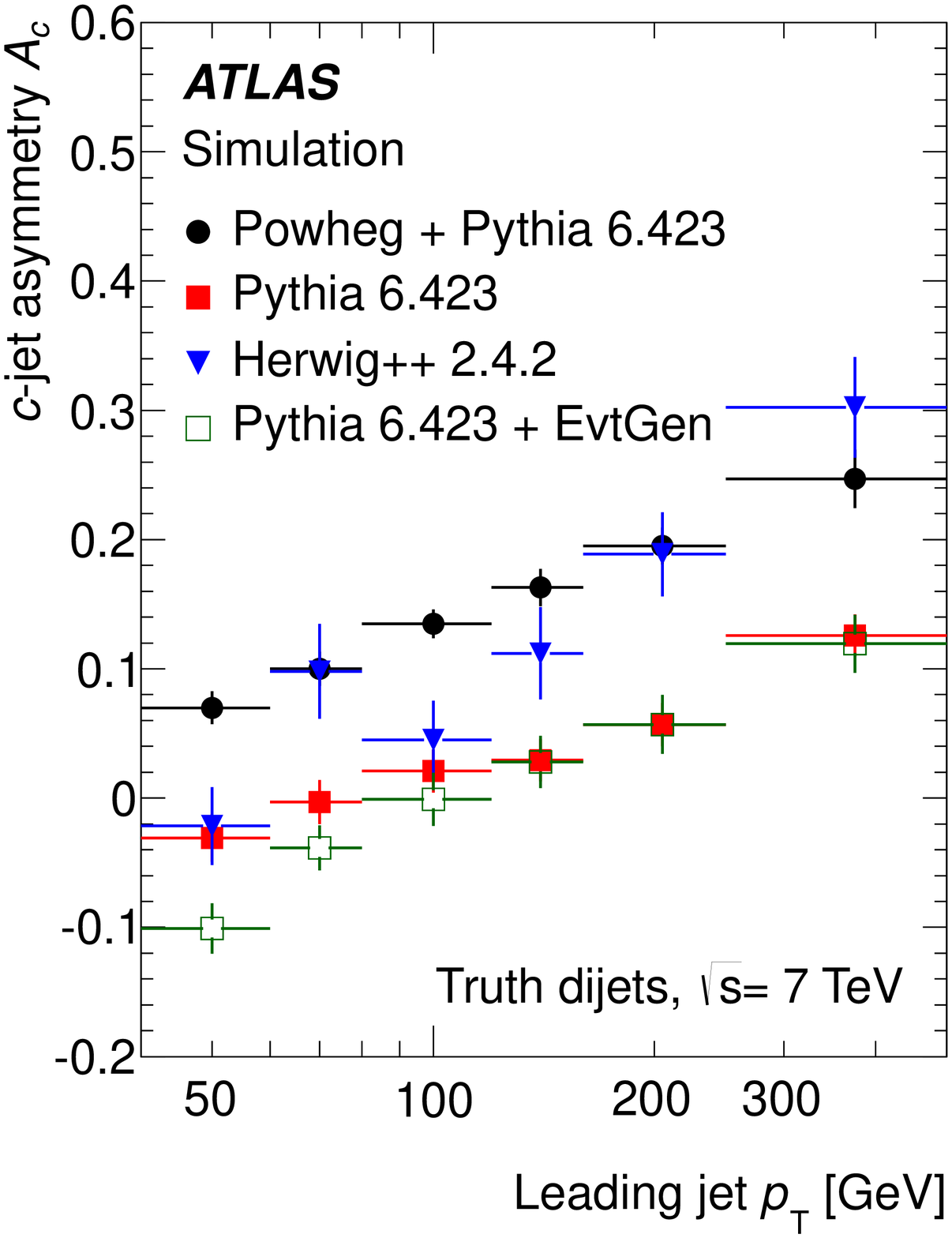}
\label{figures:asymflav2}
}
 \caption {The asymmetries in the amount of (a) bottom  and (b) charm truth particle jets 
  as taken from {\sc Powheg}+{\sc Pythia}~6.423 (black points), {\sc Pythia}~6.423 (squares), 
  Herwig++~2.4.2 (triangles) and {\sc Pythia}+{\sc EvtGen} (open squares) 
  in leading and subleading jets, for each leading jet $\pT$ bin used in the analysis.}
\label{figures:asymflav}
\end{center}
\end{figure}

The kinematic properties of the partons produced in hadro\-nic
interactions are mostly flavour independent, if mass effects are neglected.
The two back-to-back partons with the highest $\pT$ in the event should therefore not show
any significant flavour-dependent difference in their kinematic features.
However, the partons can be studied only through the corresponding jet properties after hadronisation. 
Heavy fla\-vour quark presence in a jet can influence the jet properties through the following mechanisms:
\begin{itemize}
\item Semileptonic decays of heavy flavour hadrons decrease the jet energy, because neutrinos are not detected
and the muon energy is not measured in the calorimeter.
This energy loss is absent for light jets and is very different for bottom and charm jets.
\item 
If several heavy flavour quarks appear in the jet fragmentation process (e.g. via gluon splitting) one of them
can be left outside the jet
volume by the jet reconstruction algorithm, which leads to a reduction in the jet energy.
\end{itemize}
As a result, the average jet energy for heavy flavours becomes smaller than the
jet energy for light flavours, such that heavy flavour jets are predominantly produced as subleading jets 
in the mixed-flavour dijet pairs.
This effect can be described using a flavour asymmetry defined as 
\begin{equation}\label{eq:asymmetry}
A_{b,c} = \frac{N_{b,c}^{SL}}{N_{b,c}^{L}}-1,
\end{equation}
where $N_{b,c}^{L,SL}$ denote the number of leading or subleading bottom or charm jets.
The predictions for $A_{b,c}$ given by different Monte Carlo generators are shown in  Fig.~\ref{figures:asymflav} 
for the truth particle jets defined in Sect.~\ref{sec:Mc}.

\fussy

{\sc Powheg}, which includes higher-order QCD effects, predicts a
significant flavour asymmetry which increases strong\-ly with jet $\pT$. 
The flavour asymmetry predictions of the LO {\sc Pythia} generator are smaller than those of the NLO
{\sc Powheg} generator.
The latter uses {\sc Pythia}~6.423 for the fragmentation and 
thus shares the same description of the decays of heavy flavour hadrons. 
Since the influence of the different parton distribution functions was also found to be negligible,
the differences in $A_{b,c}$ between these
generators (Fig.~\ref{figures:asymflav}) should be attributed primarily to NLO QCD effects.
The LO Herwig++ generator employs another fragmentation model and predicts asymmetries similar to the {\sc Powheg} ones,
although with a somewhat different $\pT$ dependence.

\sloppy

For the measurement of the dijet flavour fractions, this flavour asymmetry needs to be correctly described in the data analysis. 
The fact that the Monte Carlo generators predict significantly
different asymmetries indicates that $A_{b,c}$ should be determined
directly from the data.

\section{Secondary vertex reconstruction and analysis templates}
\label{sec:SVreco}
Secondary vertices are displaced from the primary vertex because they originate from the decays of long-lived particles.
Kinematic properties of these vertices, e.g. the invariant mass or total energy of the outgoing particles, depend on the corresponding properties 
of the original heavy flavour hadrons and are therefore different for bottom and charm jets. 
Reconstructed secondary vertices in light jets are mainly due to
$K^0_S$ and $\Lambda$~\cite{Nakamura:2010zzi} decays, interactions in
the detector material, or fake vertices. The fake reconstructed vertices are composed of tracks 
which occasionally get close together due to a high density of tracks in the jet core and track reconstruction errors.
Their properties are very different from those of heavy flavour decays. 
The current analysis exploits these differences by combining the kinematic features of the reconstructed secondary vertices 
in an optimal way into templates for bottom, charm and light jets.

\subsection{Secondary vertex reconstruction in jets}

The vertex reconstruction algorithm aims at a high reconstruction efficiency and therefore determines vertices
in an inclusive way, i.e. a single secondary vertex is
fitted for each jet. In the case of a bottom hadron decay, 
the subsequent charm hadron decay vertex is usually close to the bottom one and is therefore not reconstructed separately. 
A detailed discussion of the algorithm and its performance can be found in the $b$-tagging chapter of Ref.\cite{atlas_performance}. 
The reconstruction starts by 
combining pairs of good quality tracks inside jets to make vertices,
where the latter are required to be displaced significantly from the primary interaction vertex.
The two-track vertices coming from $K^0_S$ and $\Lambda$ decays and interactions in the detector material are removed from further consideration.
For the light jets, the remaining candidates after this cleaning are mainly fake vertices.
All remaining two-track vertices are merged into a single vertex. 
This vertex is refitted iteratively by removing tracks until a good vertex fit quality is obtained. 
The corresponding decay length is defined as a signed quantity, where the sign is fixed by the projection 
of the decay length vector---the vector pointing from the primary event vertex to the secondary vertex---onto the jet axis. 
The vertex is required to have a positive decay length and a total invariant mass, 
calculated using the momenta of associated particles and assigning them pion masses~\cite{Nakamura:2010zzi}, greater than $0.4\GeV$.

\subsection{Secondary vertex reconstruction efficiencies}
\label{subsec:SVRecoEff}

\begin{figure*}[!hbt]
\begin{center}
\subfigure[]{
\includegraphics[width=0.315\textwidth]{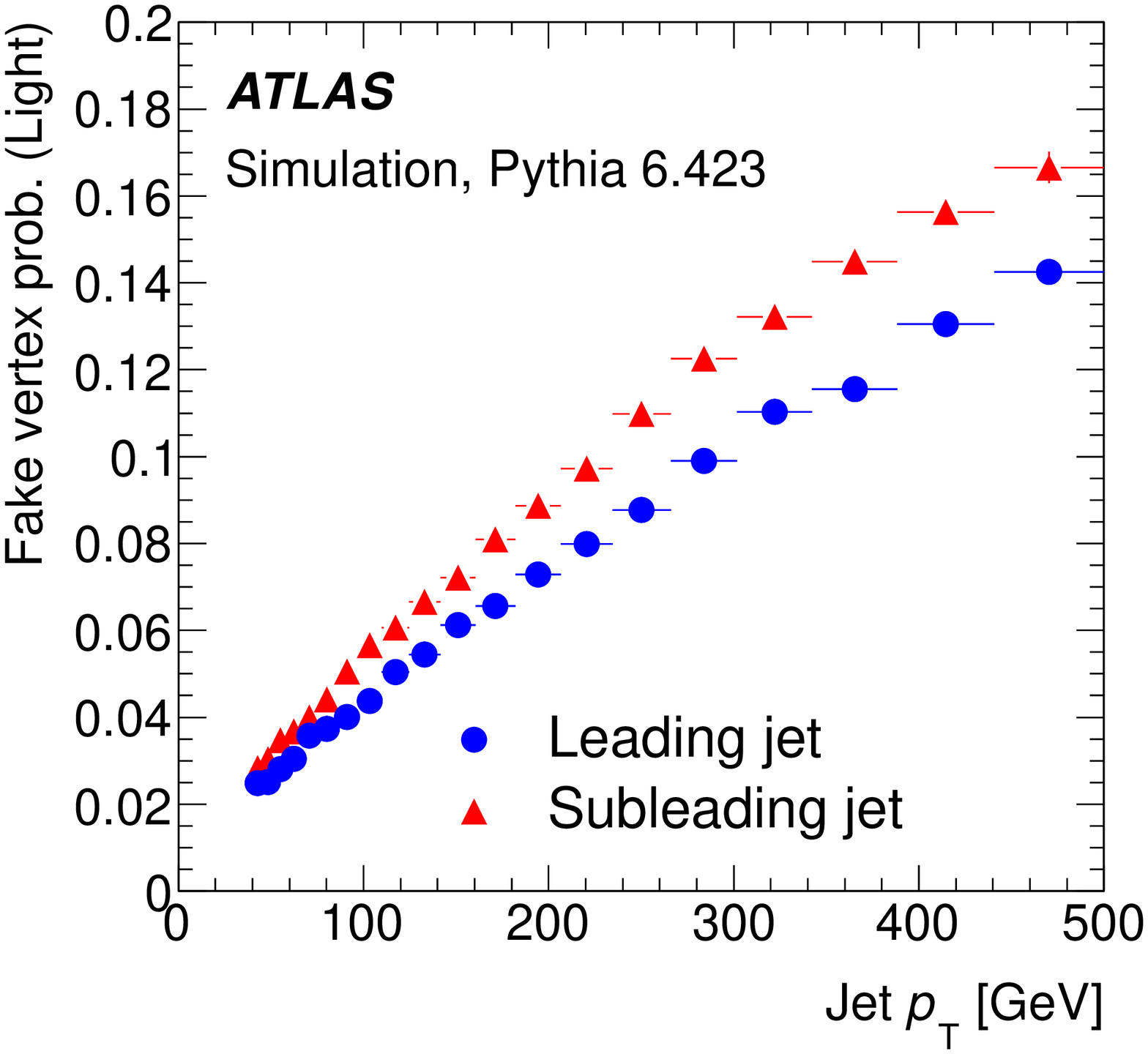}
\label{figures:SVRecoEffa}
}
\subfigure[]{
\includegraphics[width=0.315\textwidth]{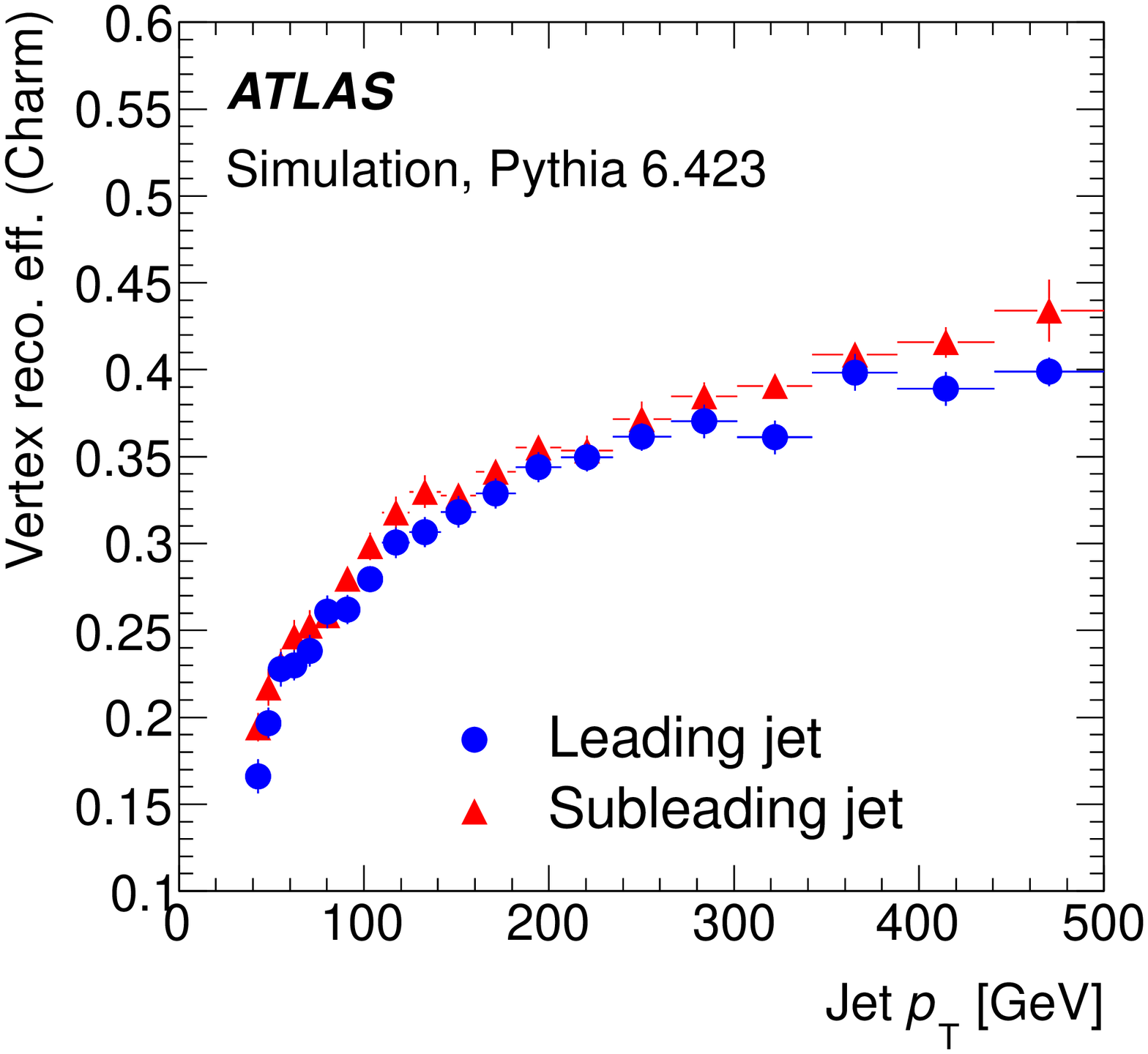}
\label{figures:SVRecoEffb}
}
\subfigure[]{
\includegraphics[width=0.315\textwidth]{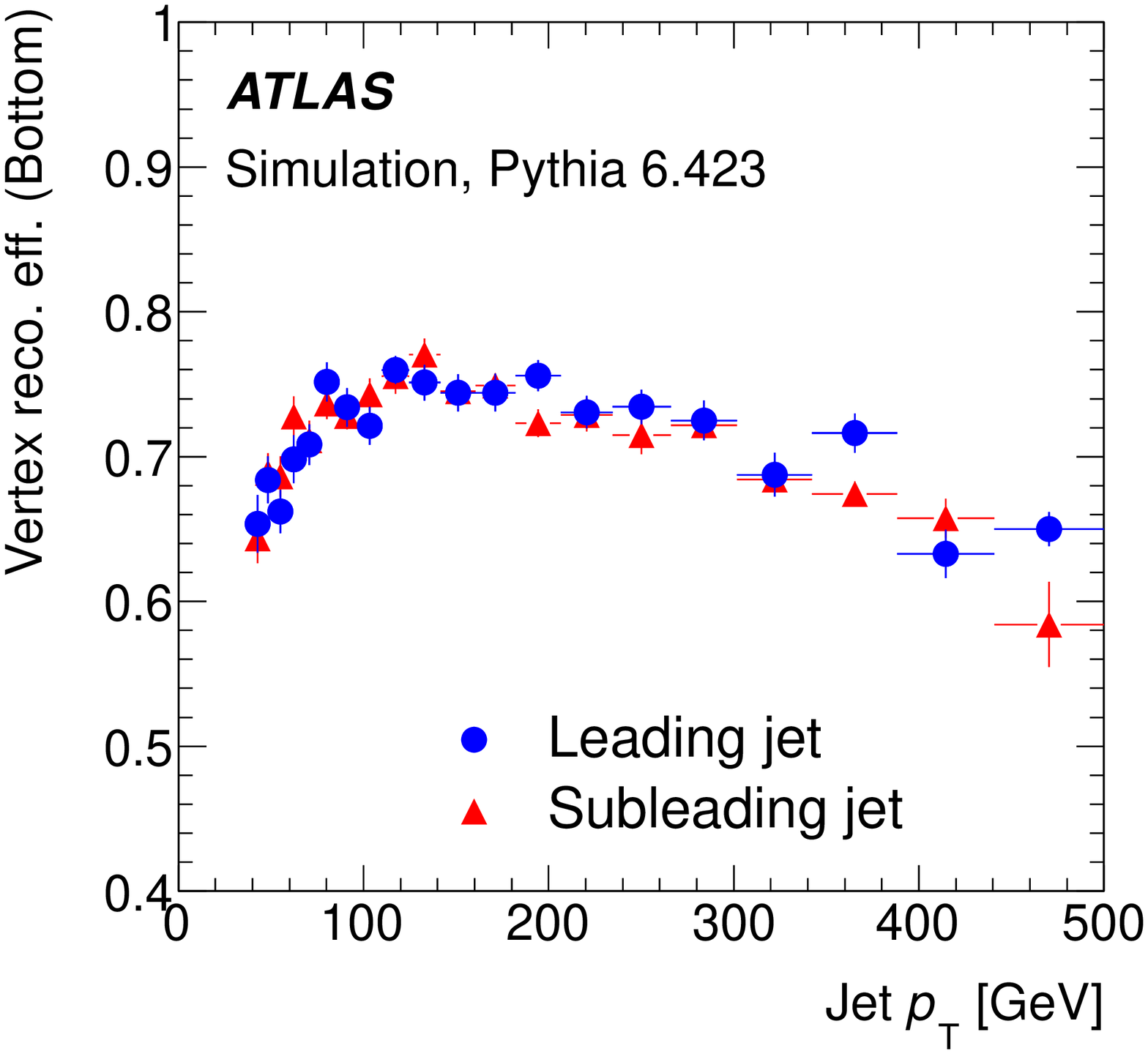}
\label{figures:SVRecoEffc}
}
 \caption {The reconstruction probabilities for fake vertices in (a) light jets, as well as the reconstruction efficiencies 
  for secondary vertices in (b) charm and (c) bottom jets, are displayed as a function of the jet $\pT$ as predicted by {\sc Pythia}~6.423.}
\label{figures:SVRecoEff}
\end{center}
\end{figure*}

The secondary vertex reconstruction efficiency is dependent on the jet
$\pT$ due to several effects such as the $\pT$ dependence of the track 
reconstruction accuracy and the increase of the flight distance of heavy flavour hadrons with growing jet $\pT$.
The probability of reconstructing a fake vertex in a light jet is also
affected by the increase of the number of tracks in a jet with jet $\pT$.
Due to the $\pT$-dependent vertex efficiency and different $\pT$
distributions for leading and subleading jets in dijet pairs, the number of
reconstructed secondary vertices in these jets are different.

\fussy

 The secondary vertex reconstruction efficiencies 
predicted by the ATLAS detector simulation based on dijet events from {\sc Pythia}~6.423 are shown in Fig.~\ref{figures:SVRecoEff}.
There is no difference between secondary vertex reconstruction efficiencies in leading and subleading jets for charm and bottom jets.
However, the fake vertex reconstruction probability in light jets is noticeably higher for subleading jets. 
This requires the introduction of two separate secondary vertex probabilities
for leading and subleading light jets.
\sloppy

\subsection{Template construction and features}
\label{subsec:TemplateFeatures}

\begin{figure*}[!hbt]
\begin{center}
\subfigure[]{
\includegraphics[width=0.315\textwidth]{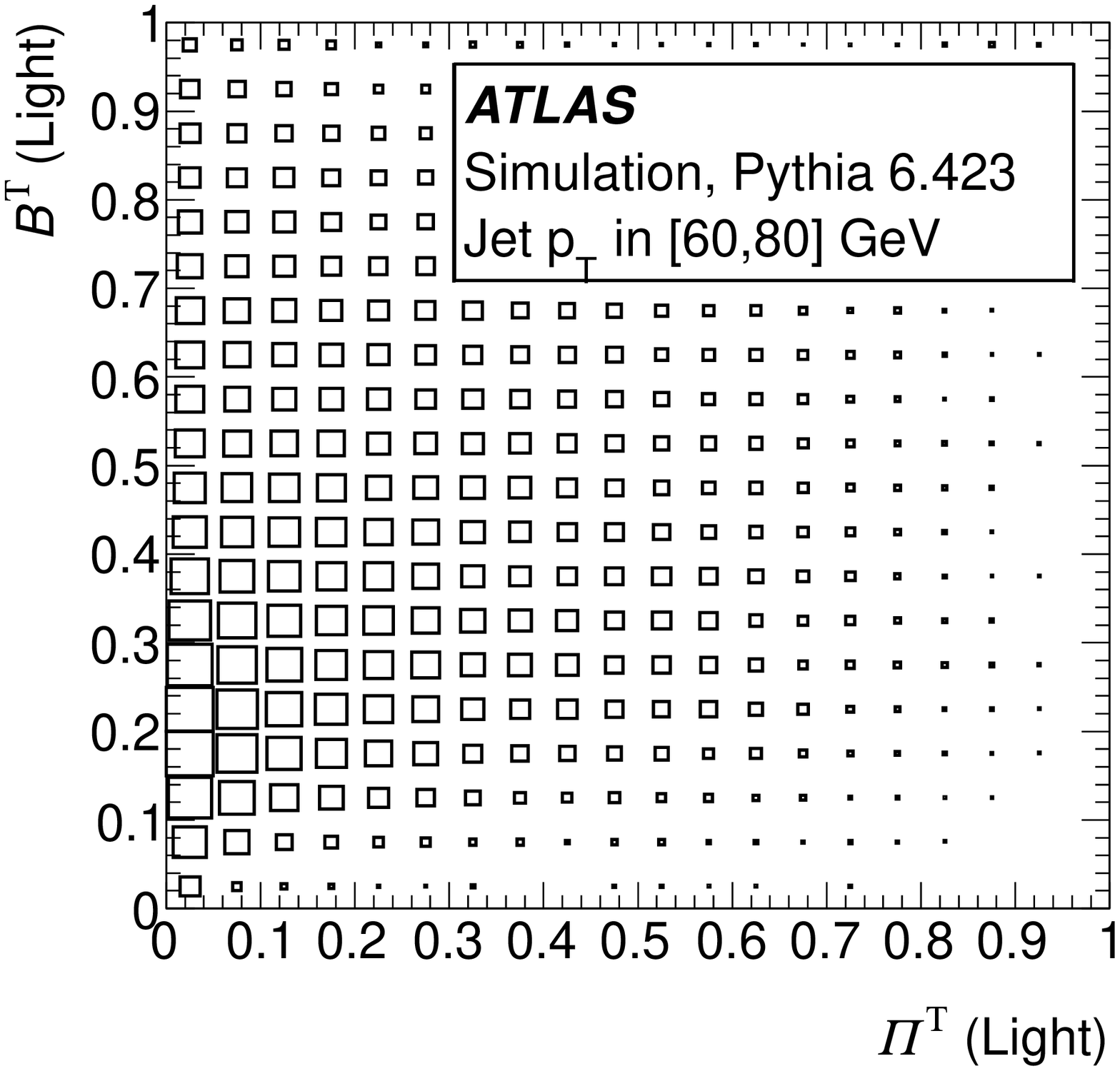}
\label{figures:FitVar2Da}
}
\subfigure[]{
\includegraphics[width=0.315\textwidth]{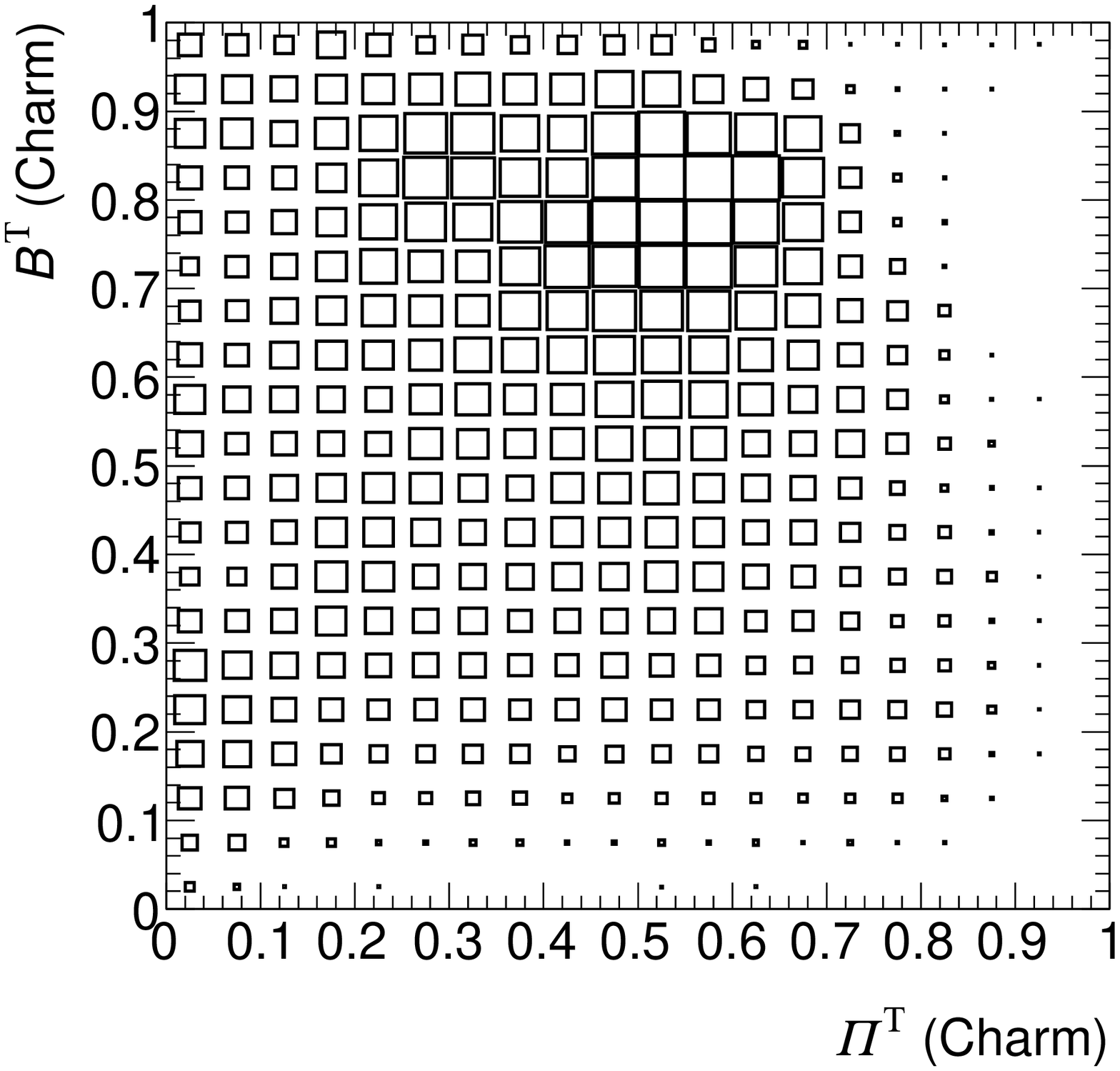}
\label{figures:FitVar2Db}
}
\subfigure[]{
\includegraphics[width=0.315\textwidth]{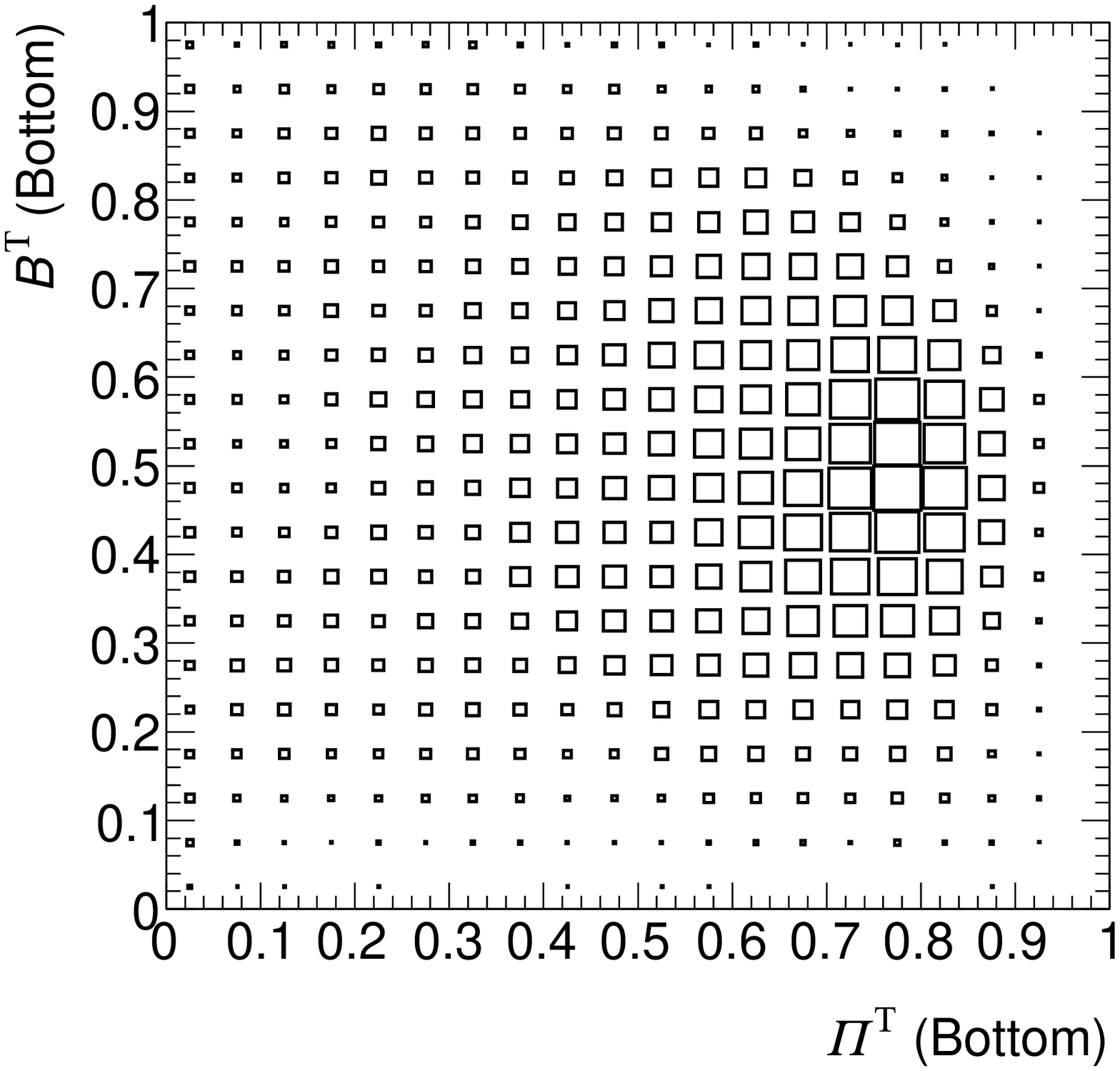}
\label{figures:FitVar2Dc}
}
 \caption {Two-dimensional distributions of $\Pi^{\top}$ and $B^{\top}$ (flavour templates) obtained with {\sc Pythia}~6.423 for (a) light, (b) charm and (c) bottom jets with $\pt$ in the bin $[60,80]\GeV$.}
\label{figures:FitVar2D}
\end{center}
\end{figure*}

The specific choice of the kinematic variables for the dijet flavour measurement is driven by the requirement 
to have maximal sensitivity to the flavour content. 
Furthermore, if several variables are to be used, the correlations between them should be kept small. 
Another important requirement is a minimal dependence on the jet $\pT$ and rapidity, in order to minimise systematic effects due to a possible
$\pT$ or rapidity mismatch between data and Monte Carlo simulation. Also,
$\pT$-invariant variables allow a robust analysis to be made over a wide range of $\pt$.

For this study the following two variables are chosen:
\begin{equation}\label{pvar}
\Pi=\frac{m_{\,\mathrm{vertex}}-0.4\GeV}{m_B} \cdot \frac{\sum\limits_{\mathrm{vertex}}E_i}{\sum\limits_{\mathrm{jet}}E_i}
\end{equation}
\begin{equation}\label{bvar}
B=\frac{\sqrt{m_B} \cdot \sum\limits_{\mathrm{vertex}}|\overrightarrow {\pT}_i|}{m_{\,\mathrm{vertex}} \cdot \sqrt{{\pT}_{jet}}},
\end{equation}
where each sum indicates whether the summation is performed over particles associated with the secondary vertex, or over all charged particles in the jet. 
Particle transverse momentum and energy are denoted as $\pT$ and $E$, respectively.
In essence, $\Pi$ is the product of the invariant mass of the particles associated with the vertex ($m_{\,\mathrm{vertex}}$) 
and the energy fraction of these particles with respect to all charged particles in the jet.
The $0.4\GeV$ constant in Eq.(\ref{pvar}) is the cut value used for the secondary vertex selection in this analysis. 
The parameter $B$ corresponds approximately to the relativistic $\gamma$ factor of the system composed of the particles associated with the vertex, 
normalised to the square root of the jet transverse momentum. 
The $m_B=5.2794\GeV$ constant is the average $B$-meson mass~\cite{Nakamura:2010zzi} and is used for normalisation.

To facilitate the fit procedure, 
the variables are transformed into the interval [0,1]:
\begin{equation}\label{pvart}
\Pi^{\top}=\frac{\Pi}{\Pi + 0.04}
\end{equation}
\begin{equation}\label{bvart}
B^{\top}=\frac{B\cdot B}{B \cdot B + 10.}.
\end{equation}
The tuning constants $0.04$ in Eq.(\ref{pvart}) and $10$ in Eq.(\ref{bvart}) have been chosen to maximise 
the difference in the mean values between the light and heavy flavour distributions.

\sloppy

\begin{figure*}[!hbt]
\begin{center}
\subfigure[Light jets in {[40,60]} GeV]{
\includegraphics[width=0.48\textwidth]{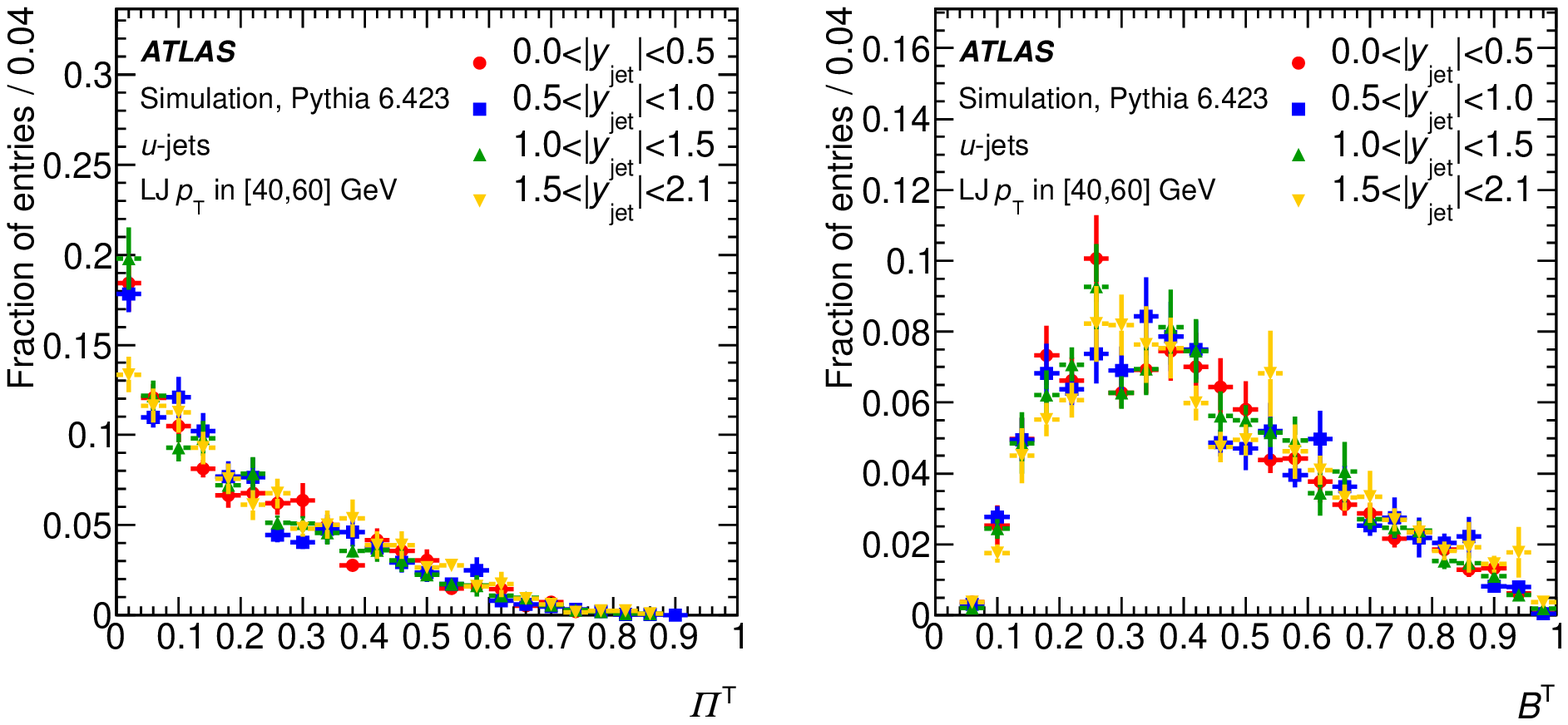}
\label{figures:LJTemplatesA}
}
\subfigure[Light jets in {[250,500]} GeV]{
\includegraphics[width=0.48\textwidth]{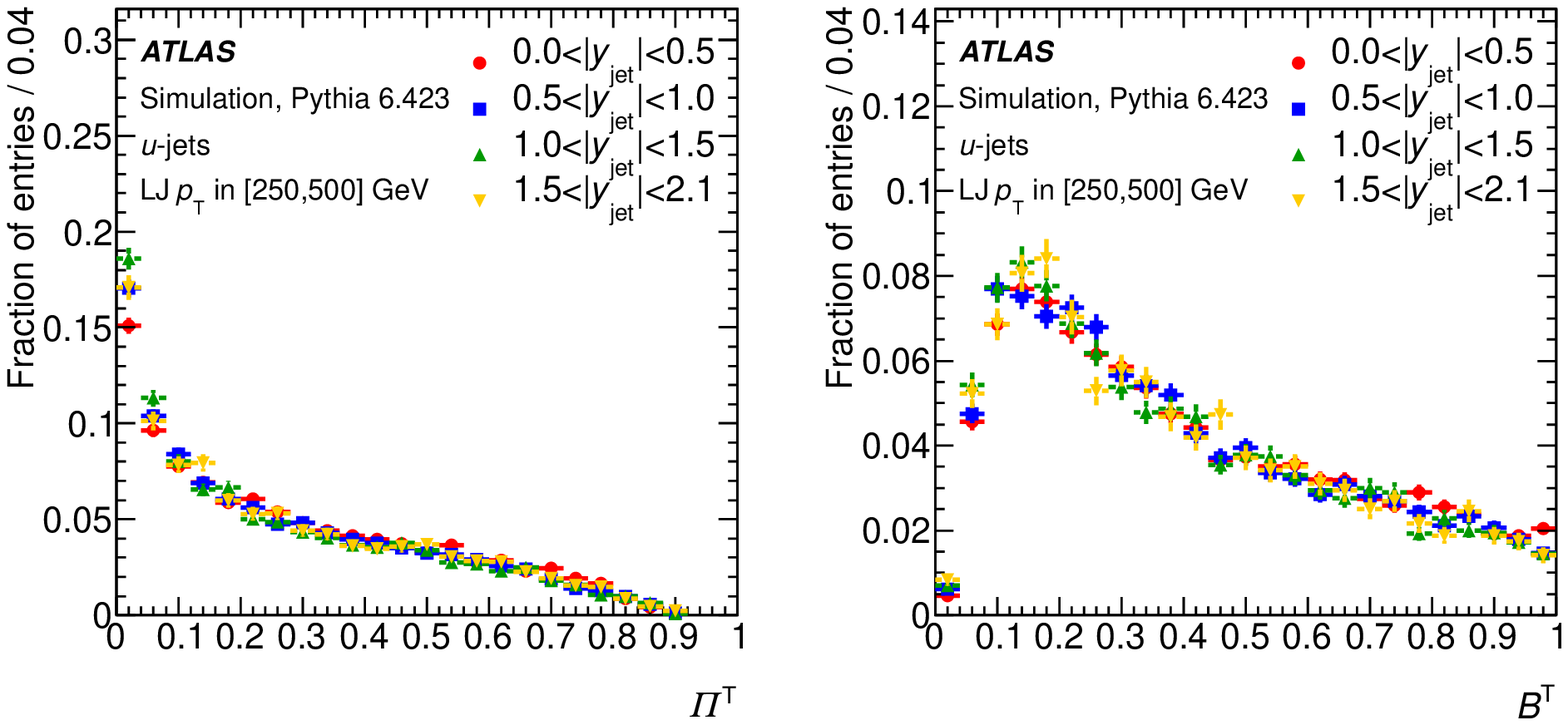}
\label{figures:LJTemplatesB}
}
 \caption{ The $\Pi^{\top}$  and $B^{\top}$ distributions of light jets in the \subref{figures:LJTemplatesA} $[40,60]\GeV$ and \subref{figures:LJTemplatesB} $[250,500]\GeV$
   leading jet (LJ) $\pT$ analysis bins obtained with fully simulated {\sc Pythia}~6.423 dijet events. 
   The distributions are shown in different jet rapidity ranges. } 
\label{figures:LJTemplates}
\end{center}
\end{figure*}

\begin{figure*}[!hbt]
\begin{center}
\subfigure[Charm jets in {[40,60]} GeV]{
\includegraphics[width=0.48\textwidth]{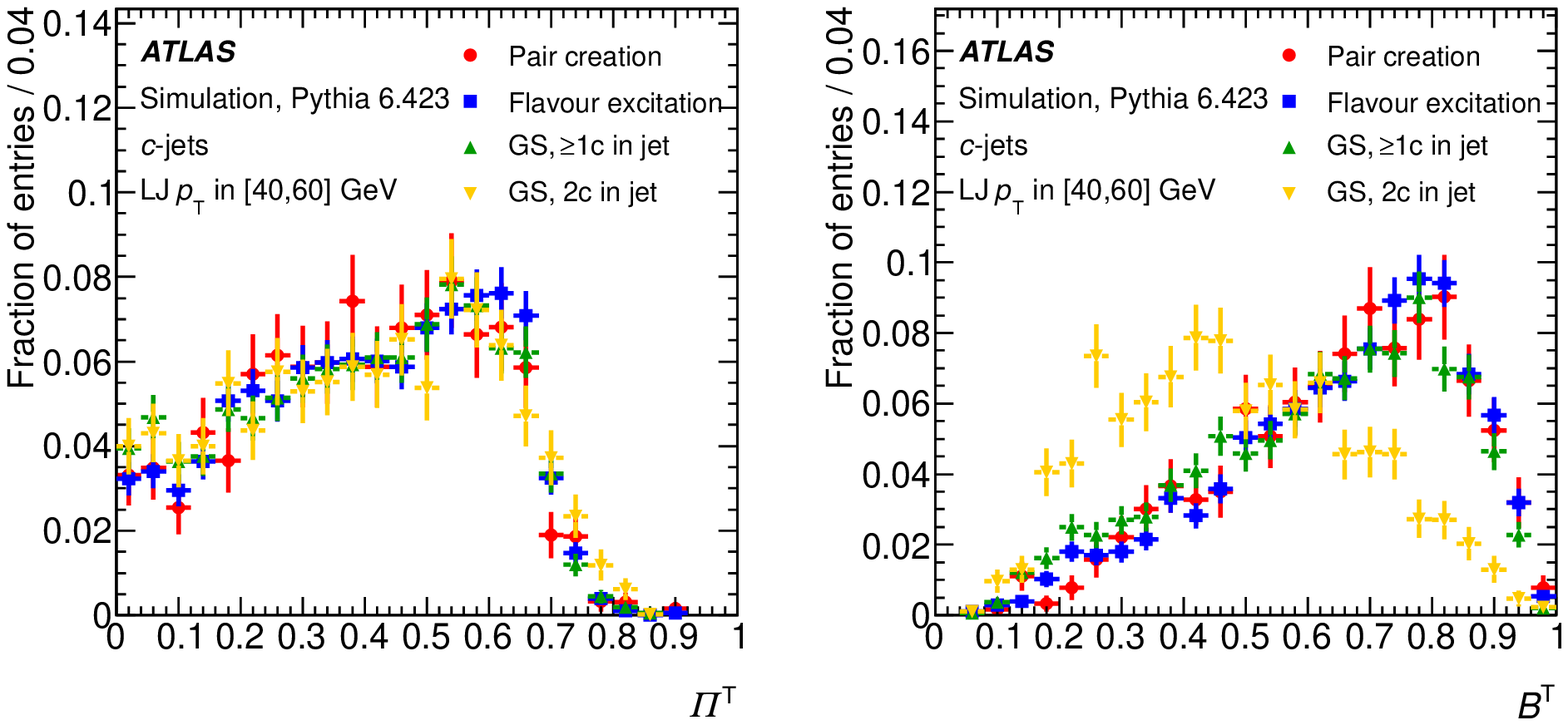}
\label{figures:CJTemplatesA}
}
\subfigure[Charm jets in {[250,500]} GeV]{
\includegraphics[width=0.48\textwidth]{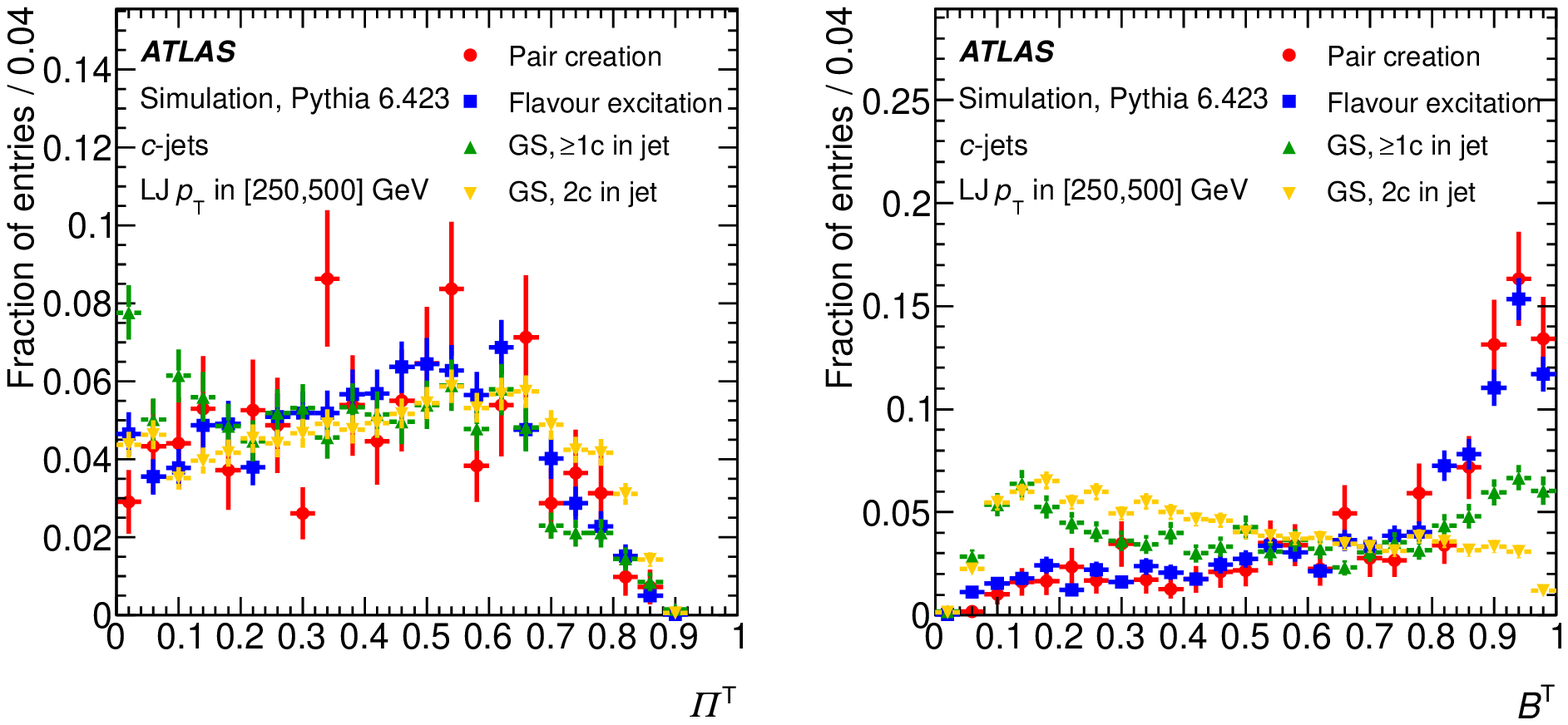}
\label{figures:CJTemplatesB}
}
\subfigure[Bottom jets in {[40,60]} GeV]{
\includegraphics[width=0.48\textwidth]{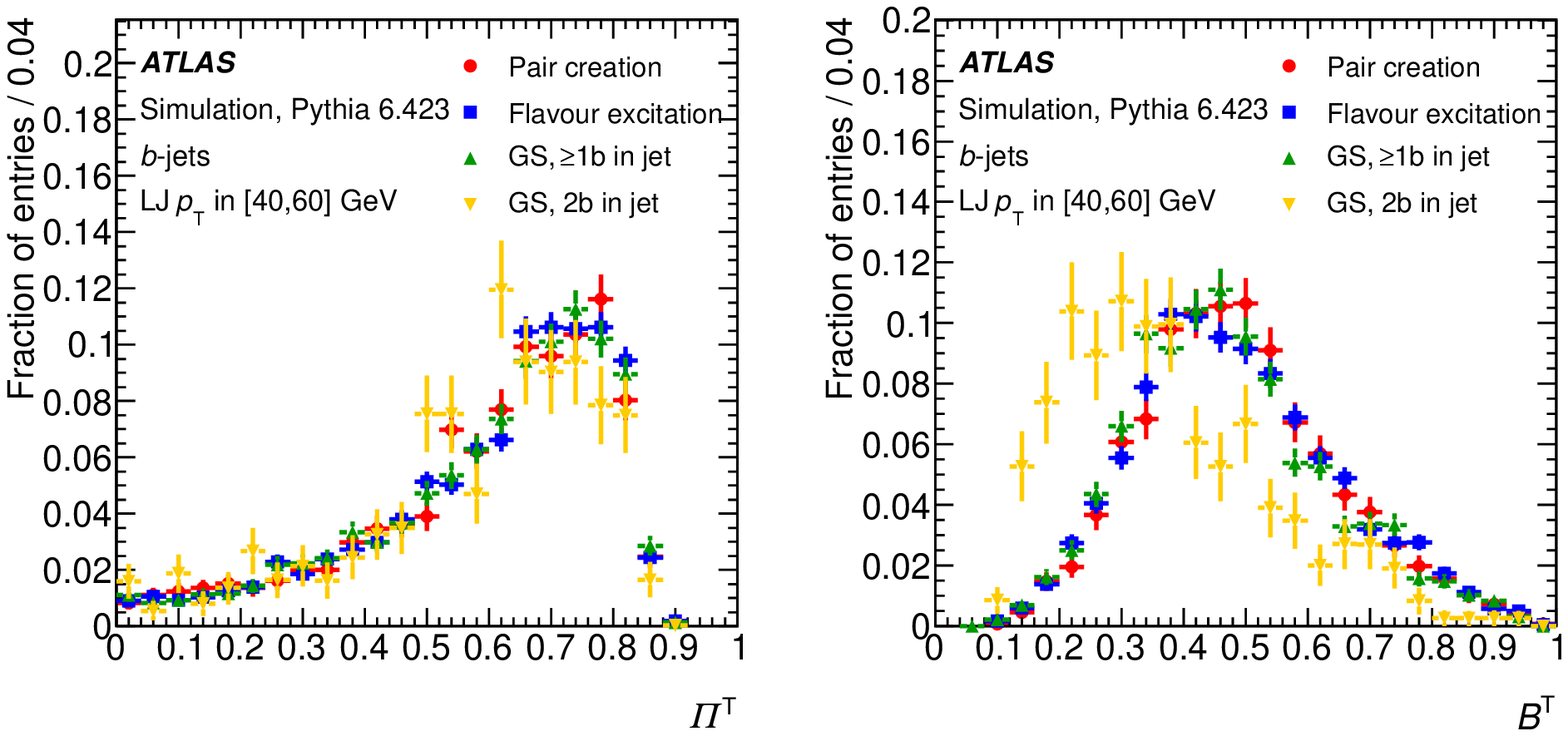}
\label{figures:BJTemplatesA}
}
\subfigure[Bottom jets in {[250,500]} GeV]{
\includegraphics[width=0.48\textwidth]{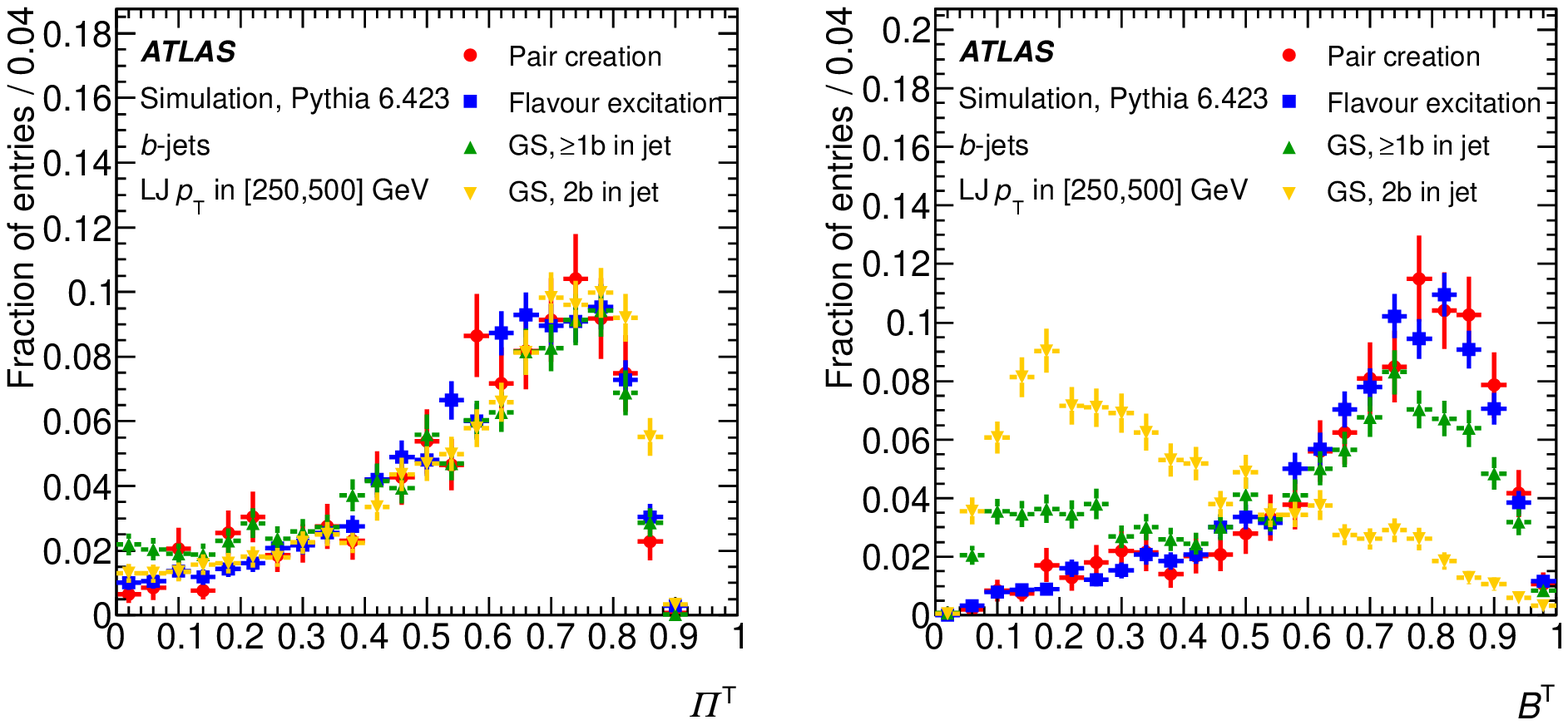}
\label{figures:BJTemplatesB}
}
 \caption{ The $\Pi^{\top}$  and $B^{\top}$ distributions in the
   $[40,60]\GeV$ leading jet (LJ) $\pT$ range for \subref{figures:CJTemplatesA} charm jets and
   \subref{figures:BJTemplatesA} bottom jets as well as in the
   $[250,500]\GeV$ range for \subref{figures:CJTemplatesB} charm jets
   and \subref{figures:BJTemplatesB} bottom jets  obtained with fully simulated {\sc Pythia}~6.423 dijet events. 
   The distributions are shown separately for jets stemming from quark pair creation, 
   heavy flavour quark excitation, gluon splitting (GS) with one or two heavy flavour quarks inside the jet.
   All distributions are normalised separately to unit area.
   }
\label{figures:CBJTemplates}
\end{center}
\end{figure*}

Joint distributions of these observables  
are shown in Fig.~\ref{figures:FitVar2D} for light, charm and bottom jets in the $[60,80]\GeV$ bin,
as predicted by the full detector simulation of {\sc Pythia}~6.423 events.
These two-dimensional distributions are used as flavour templates $U(\Pi^{\top},B^{\top})$, 
$C(\Pi^{\top},B^{\top})$ and $B(\Pi^{\top},B^{\top})$ in the analysis as detailed in Section \ref{sec:Method}.
Features of the observables are also illustrated 
in Figs.~\ref{figures:LJTemplates} and \ref{figures:CBJTemplates}.
Both $\Pi^{\top}$ and $B^{\top}$ are independent of jet rapidity for all jet flavours.
This is illustrated in Fig.~\ref{figures:LJTemplates} for the light jet templates, 
which are most sensitive to reconstruction and detector effects.
The $\Pi^{\top}$ variable is very similar in shape in the $[40,60]\GeV$ and $[250,500]\GeV$ bins and 
is only weakly $\pT$-dependent. Figure~\ref{figures:CBJTemplates} demonstrates that $\Pi^{\top}$ is 
only weakly dependent on the different heavy flavour production mechanisms described in Sect.~\ref{sec:Predictions}.
In contrast, the $B^{\top}$ variable is sensitive to the gluon splitting contribution, in particular to the case where 
this mechanism produces two quarks of the same flavour in a jet. In addition $B^{\top}$ has a distinct $\pT$ dependence.
However, the $B^{\top}$ variable provides good sensitivity to the charm contribution. 
No difference in flavour templates between leading and subleading jets is observed.

\fussy

The fraction of jets with two heavy quarks produced in gluon splitting 
may be incorrectly predicted by the {\sc Pythia} simulation, especially in the high $\pT$ region
where this contribution becomes large (see Fig.~\ref{figures:BProd}).
This phenomenon was discussed in more detail in~\cite{BShapeCDF}.
Therefore a separate contribution of doubly-flavoured jets is included in the analysis,
to account for the corresponding dependence of the $B^{\top}$ variable. 
The two-dimensional template for bottom jets is replaced by the two-component template
\begin{equation}\label{datadrivenb}
 B(\Pi^{\top},B^{\top}) \rightarrow (1-b_2)\cdot B(\Pi^{\top},B^{\top}) + b_2 \cdot B_2(\Pi^{\top},B^{\top}),
\end{equation}
where $B_2(\Pi^{\top},B^{\top})$ is a template for jets with two $b$-hadrons and $b_2$ is a parameter governing the deviation from the default 2b-jet $B(\Pi^{\top},B^{\top})$ content provided by {\sc Pythia}~6.423.
The charm jet template is modified similarly with substitutions $b_2 \rightarrow c_2$ and 
$B_2(\Pi^{\top},B^{\top}) \rightarrow C_2(\Pi^{\top},B^{\top})$.
Using Eq.(\ref{datadrivenb}), the heavy flavour template shapes can be obtained directly from the data by optimising the $b_2$ and $c_2$ parameters 
to achieve the best possible data description. As is demonstrated in Sect.~\ref{sec:Results}, 
the adjustment of the contribution of jets with two $b$-hadrons to the
bottom template significantly improves the overall quality of the
description of the dijet data.

\subsection{Template tuning on data using track impact parameters}
\label{subsec:TemplateTune}

\sloppy
The secondary vertex reconstruction algorithm uses track impact parameters divided by their measurement uncertainties for the vertex search,
thus its results depend crucially on the track impact parameter resolution.
A good description of the track impact parameter accuracy and the corresponding covariance matrix
is therefore mandatory in the detector simulation, in order for the
secondary vertex templates to be constructed correctly.

 To improve the agreement between data and Monte Carlo simulation, the analysis templates are tuned on data.
Firstly, an additional track impact parameter smearing is applied to the {\sc Pythia} events. 
To estimate the necessary amount of smearing, the data and Monte Carlo track impact parameter distributions
are compared in bins of track $\pT$ and pseudorapidity~\cite{ATLAS-CONF-2010-070}.
However, the smearing procedure does not correct the track covariance matrices. 
A second step is therefore taken. Two sets of templates are produced,
using both the smeared and non-smeared {\sc Pythia}~6.423 samples.
A normalised mixture is then compared with the data, using secondary vertices with negative decay length to obtain the optimal mixing fraction.
These vertices depend only weakly on the exact flavour content of jets and are not used in the dijet analysis. 
The mixing fraction is chosen to be flavour independent.
The optimal description of the data for the full $\pT$ range is obtained with a fraction $F_{\mathrm{smear}}=0.654 \pm 0.023$ for the smeared template in the mixture.
This template tuning procedure gives a significant improvement in the data fit quality in the signal region.

\fussy

\section{Analysis method}
\label{sec:Method}
\subsection{Dijet system description}

\sloppy

The secondary vertex reconstruction procedure can find vertices
with probabilities $v_U$, $v_C$ and $v_B$ for light, charm and bottom jets, respectively.
For simplicity, the $\pT$-dependence of these probabilities and the
differences between leading and subleading jets (see Sect.~\ref{subsec:SVRecoEff}) 
are neglected for the moment.
In the leading and subleading jet of a dijet event, zero, one or two secondary vertices can be reconstructed overall. 
The numbers of 2-, 1-, or 0-vertex dijet events can be calculated as:

\noindent \parbox{\columnwidth}{\begin{eqnarray} \label{2vfrac}
 \frac{N_{\rm 2V}}{N}= && v_U v_U f_{UU} + v_C v_C f_{CC} +v_B v_B f_{BB} \\
                   && {}+v_U v_C f_{CU} +v_U v_B f_{BU} +v_C v_B f_{BC} \nonumber
\end{eqnarray} }

\noindent \parbox{\columnwidth}{ \begin{eqnarray} \label{1vfrac}
 \frac{N_{\rm 1V}}{N}= & 2\left(1-v_U\right) \cdot v_U \cdot f_{UU} + 2\left(1-v_C\right) \cdot v_C \cdot f_{CC} \\
           &       + 2\left( 1-v_B\right) \cdot v_B \cdot f_{BB} \nonumber\\
           &       + \left( (1-v_U) \cdot  v_C + v_U \cdot (1-v_C) \right) \cdot f_{CU} \nonumber\\
           &       + \left( (1-v_U) \cdot  v_B + v_U \cdot (1-v_B) \right) \cdot f_{BU} \nonumber\\
           &       + \left( (1-v_C) \cdot  v_B + v_C \cdot (1-v_B) \right) \cdot f_{BC} \nonumber
\end{eqnarray} }

\begin{equation}
 N_{\rm 0V}= N - N_{\rm 1V} - N_{\rm 2V}.
\end{equation}

Here $N$ is the total number of dijet events and $f_{XX}$ is the fraction of the respective dijet flavour component chosen such that
\begin{equation}\label{depfrac} 
   f_{UU} + f_{CC} +f_{BB} +f_{CU} + f_{BU} + f_{BC} = 1.
\end{equation}

The joint distribution of the $\Pi^{\top}$ and $B^{\top}$ variables for dijet events 
with one reconstructed secondary vertex can be obtained using Eq.(\ref{1vfrac}):

\noindent\parbox{\columnwidth}{
\begin{eqnarray} \label{1vtempl}
 \lefteqn{{\cal D}(\Pi^{\top},B^{\top}) =   2 \left(1-v_U\right) v_U  f_{UU}  U(\Pi^{\top},B^{\top}) }\\
          &+ 2 \left(1-v_C\right)  v_C  f_{CC}  C(\Pi^{\top},B^{\top}) \nonumber\\
	  &+ 2 \left(1-v_B\right)  v_B  f_{BB}  B(\Pi^{\top},B^{\top}) \nonumber\\
        &+ \left\{ (1-v_U) v_C  C(\Pi^{\top},B^{\top}) + v_U (1-v_C)  U(\Pi^{\top},B^{\top}) \right\}  f_{CU} \nonumber\\
        &+ \left\{ (1-v_U) v_B  B(\Pi^{\top},B^{\top}) + v_U (1-v_B)  U(\Pi^{\top},B^{\top}) \right\}  f_{BU} \nonumber\\
        &+ \left\{ (1-v_C) v_B  B(\Pi^{\top},B^{\top}) + v_C (1-v_B)  C(\Pi^{\top},B^{\top}) \right\}  f_{BC}. \nonumber
\end{eqnarray}
}

Here ${\cal D}(\Pi^{\top},B^{\top})$ is the observed data distribution and $U(\Pi^{\top},B^{\top})$, 
$C(\Pi^{\top},B^{\top})$ and $B(\Pi^{\top},B^{\top})$ are templates derived from Monte Carlo simulation with 
\begin{eqnarray} \label{TemplateNorm}
\int U(\Pi^{\top},B^{\top})d\Pi^{\top}dB^{\top}&=&\int C(\Pi^{\top},B^{\top})d\Pi^{\top}dB^{\top} \\
                 &=&\int B(\Pi^{\top},B^{\top})d\Pi^{\top}dB^{\top} = 1. \nonumber
\end{eqnarray}

The case of two reconstructed vertices requires more careful consideration. Assuming that the two jets are independent,
the joint distribution of $\Pi^{\top}$ and $B^{\top}$ can be written considering Eq.(\ref{2vfrac}) in the following way: 

\noindent\parbox{\columnwidth}{ 
\begin{eqnarray} \label{2vtempl}
\lefteqn{{\cal D}(\Pi^{\top}_1, B^{\top}_1, \Pi^{\top}_2, B^{\top}_2) = } \\
              &\phantom{+} v_U v_U f_{UU} \cdot U(\Pi^{\top}_1,B^{\top}_1) U(\Pi^{\top}_2,B^{\top}_2) \nonumber\\ 
              &+ v_C v_C f_{CC} \cdot C(\Pi^{\top}_1,B^{\top}_1) C(\Pi^{\top}_2,B^{\top}_2) \nonumber\\
              &+ v_B v_B f_{BB} \cdot B(\Pi^{\top}_1,B^{\top}_1) B(\Pi^{\top}_2,B^{\top}_2) \nonumber\\
	\lefteqn{+ 0.5 \cdot v_U v_C f_{CU}}  \nonumber \\
	 & \times  \left\{ U(\Pi^{\top}_1,B^{\top}_1) C(\Pi^{\top}_2,B^{\top}_2)+ U(\Pi^{\top}_2,B^{\top}_2) C(\Pi^{\top}_1,B^{\top}_1) \right\} \nonumber \\
	\lefteqn{+ 0.5 \cdot v_U v_B f_{BU}}  \nonumber \\
	 & \times  \left\{ U(\Pi^{\top}_1,B^{\top}_1) B(\Pi^{\top}_2,B^{\top}_2)+ U(\Pi^{\top}_2,B^{\top}_2) B(\Pi^{\top}_1,B^{\top}_1) \right\} \nonumber \\
	\lefteqn{+ 0.5 \cdot v_C v_B f_{BC}}  \nonumber \\
	 & \times  \left\{ C(\Pi^{\top}_1,B^{\top}_1) B(\Pi^{\top}_2,B^{\top}_2)+ C(\Pi^{\top}_2,B^{\top}_2) B(\Pi^{\top}_1,B^{\top}_1) \right\} .\nonumber
\end{eqnarray}
} 
   
Provided that the templates $U(\Pi^{\top},B^{\top})$, $C(\Pi^{\top},B^{\top})$ and $B(\Pi^{\top},B^{\top})$ are given, the eight variables
$v_U,\; v_C,\; v_B,\; f_{CC},\; f_{BB},\; f_{BU},\; f_{CU},\; f_{BU}$
fully describe the properties of secondary vertices in ideal dijet events without kinematic dependencies.
Note that only five fractions are needed, since any of the six fractions depends on the others through Eq.(\ref{depfrac}). 
In this paper the quantity $f_{UU}$ is excluded.

The description of the dijet system must be modified to take into account the dijet flavour asymmetry (Sect.~\ref{sec:Asymmetry}).
The $BB$ and $CC$ dijet states are flavour-symmetric and thus do not require any modifications in their treatment.
The description of the $BC$ dijet fraction is also left symmetric because charm and bottom asymmetries partially compensate
each other and the fraction itself is small ($\le 0.5\,\%$).
Thus only the treatment of the $BU$ and $CU$ fractions has to be modified.
The analysis formalism is changed in the following way. The sample of dijet events with only one reconstructed secondary vertex is split into two subsamples,
according to whether the vertex is reconstructed in the leading or subleading jet.
These two subsamples are described separately, assuming different contributions of the $CU$ and $BU$ dijet fractions.
More specifically, the $f_{CU}$ and  $f_{BU}$ coefficients in Eq.(\ref{1vfrac}) and Eq.(\ref{1vtempl}) are replaced 
by pairs of coefficients $f_{CU}^L, \;f_{CU}^{SL}$ and $f_{BU}^L, \;f_{BU}^{SL}$ for leading and subleading jets, respectively. 
$L$ and $SL$ denote here whether the heavy flavour is in the leading or in the subleading jet. 
The jet flavour asymmetry of Eq.(\ref{eq:asymmetry})
can be rewritten as $A_{b,c}={f_{\{B,C\}U}^{SL}}/{f_{\{B,C\}U}^{L}}-1$. 
The new equations for events with a reconstructed secondary vertex in the leading jet can then be written:

\fussy

\noindent\parbox{\columnwidth}{ 
 \begin{eqnarray} \label{1vfracL}
 \frac{N_{\rm 1V}^L}{N^L}= & 2\cdot \left(1-v_U\right) \cdot v_U \cdot f_{UU}
                   + 2\cdot \left(1-v_C\right) \cdot v_C \cdot f_{CC} \\
	   &	   + 2\cdot \left( 1-v_B\right) \cdot v_B \cdot f_{BB} \nonumber\\
           &       + (1-v_U) \cdot  v_C \cdot f_{CU}^L + v_U \cdot (1-v_C) \cdot f_{CU}^{SL} \nonumber\\
           &       + (1-v_U) \cdot  v_B \cdot f_{BU}^L + v_U \cdot (1-v_B) \cdot f_{BU}^{SL} \nonumber\\
           &       + \left( (1-v_C) \cdot  v_B + v_C \cdot (1-v_B) \right) \cdot f_{BC} \nonumber
\end{eqnarray} }

\noindent\parbox{\columnwidth}{ 
\begin{eqnarray} \label{1vtemplL}
  \lefteqn{ {\cal D}^L(\Pi^{\top},B^{\top}) = 2\cdot \left(1-v_U\right)  v_U  f_{UU} \cdot U(\Pi^{\top},B^{\top})} \\
        &+ 2\cdot        \left(1-v_C\right)  v_C  f_{CC} \cdot C(\Pi^{\top},B^{\top}) \nonumber\\
	&+ 2\cdot      \left( 1-v_B\right)  v_B  f_{BB} \cdot B(\Pi^{\top},B^{\top}) \nonumber\\
    &+ (1-v_U)   v_C \cdot C(\Pi^{\top},B^{\top}) f_{CU}^L + v_U  (1-v_C) \cdot U(\Pi^{\top},B^{\top}) f_{CU}^{SL} \nonumber\\
    &+ (1-v_U)   v_B \cdot B(\Pi^{\top},B^{\top}) f_{BU}^L + v_U  (1-v_B) \cdot U(\Pi^{\top},B^{\top}) f_{BU}^{SL} \nonumber\\
    &+ \left\{ (1-v_C)   v_B \cdot B(\Pi^{\top},B^{\top}) + v_C  (1-v_B) \cdot C(\Pi^{\top},B^{\top}) \right\} f_{BC} \nonumber.
\end{eqnarray} }

The corresponding equations for dijet events with a reconstructed secondary vertex in the subleading jet can be obtained 
from Eq.(\ref{1vfracL}) and Eq.(\ref{1vtemplL}) by substituting $f_{CU}^L \leftrightarrow f_{CU}^{SL}$ and $f_{BU}^L \leftrightarrow f_{BU}^{SL}$.

\subsection{Data fitting function}

The complete dijet model combines all the ingredients presented in the previous sections. 
The formulae above can be modified to take into account the dependence
of the vertex reconstruction efficiencies on jet $\pT$, as well as on
whether jets are leading or subleading
(Sect.~\ref{subsec:SVRecoEff}).  Variable fractions of jets with two
bottom or charm quarks inside can also be incorporated (Sect.~\ref{subsec:TemplateFeatures}).
The full model has the following set of parameters:
\begin{eqnarray}\label{fitparfull}
v_U^L(\pT), v_U^{SL}(\pT), v_C(\pT), v_B(\pT), \\
f_{BB}, f_{BC}, f_{CC}, f_{BU}, f_{CU}, \nonumber\\
A_c, c_2, A_b, b_2. \nonumber
\end{eqnarray}

In order to reduce the set of parameters in the model to the maximum
that is affordable with the 2010 data statistics, 
additional assumptions need to be made.
The charm and bottom vertex reconstruction efficiencies are defined mainly by heavy flavour hadron lifetimes and 
heavy parton fragmentation functions, which are known well from previous experiments.
Therefore, Monte Carlo predictions for $v_B$ and $v_C$ are more robust 
than the fake vertex probability in light jets $v_U$, which is governed mainly by detector and reconstruction accuracies.
The charm asymmetry $A_c$ is smaller than the bottom one (Fig.~\ref{figures:asymflav}) and the admixture of jets with two charm quarks influences the charm template shape less than in the bottom case (Fig.~\ref{figures:CBJTemplates}).
Therefore, the following simplifications are used in the analysis: 
\begin{itemize} 
\item The fraction of jets with two charm quarks is set to the baseline {\sc Pythia}~6.423 prediction.
\item The charm jet asymmetry is fixed to $A_c=\mathrm{max}(0,A_c^{\mathrm{MC}})$ using the {\sc Pythia}~6.423 prediction, see Fig.~\ref{figures:asymflav}.
\item The $\pT$-dependent parameterisations obtained with the full ATLAS detector simulation (Fig.~\ref{figures:SVRecoEff})
      are used for bottom and charm vertex reconstruction efficiencies \\ 
 $v_C(\pT)=v_C^{\mathrm{MC}}(\pT)$ and $v_B(\pT)=v_B^{\mathrm{MC}}(\pT)$.
\item The light jet vertex reconstruction probabilities are para\-metrised as 
$v_U^L(\pT)=sv_U^L \cdot v_U^{L(\mathrm{MC})}(\pT)$ and  \\
$v_U^{SL}(\pT)=sv_U^{SL} \cdot v_U^{SL(\mathrm{MC})}(\pT)$ for leading and subleading jets, respectively.
Here $v_U^{L(\mathrm{MC})}(\pT)$ and $v_U^{L(\mathrm{MC})}(\pT)$ are the $\pT$-dependent
secondary vertex rates in light jets, obtained with the full detector simulation 
and shown in Fig.~\ref{figures:SVRecoEff}. The scaling factors $sv_U^L$, $sv_U^{SL}$ are allowed to vary in the fit.
\end{itemize}

The final model has a reduced set of nine free parameters:
\begin{equation}\label{fitpar}
sv_U^L, sv_U^{SL},f_{BB},f_{BC},f_{CC},f_{BU},f_{CU},A_b,b_2.
\end{equation}
This simplified model is used for fitting. Systematic effects
originating from the simplifications above are included in the systematic uncertainties
on the flavour fraction measurements.

\subsection{Validation of the analysis method}
\label{sec:MethodFastMC}

A dedicated simulation technique was developed to validate the analysis method.
It uses a set of secondary vertices, which are reconstructed in all jets
in the dijet sample generated with {\sc Pythia} after full ATLAS simulation, and are stored in a dedicated database in bins of jet $\pT$, rapidity and flavour.

To produce a dijet event, the $\pT$ and $|y|$ values for each jet are generated randomly according to the corresponding data distributions.
Jet flavours are assigned according to the predefined dijet flavour fractions and the flavour asymmetries (Sect.~\ref{sec:Asymmetry}).
The flavour-dependent vertex reconstruction efficiencies
(Fig.~\ref{figures:SVRecoEff}) determine whether a secondary vertex is reconstructed in the generated jet.
The vertex parameters are then taken from a fully simulated secondary
vertex, picked at random from the vertex database bin with corresponding $\pT$ and $|y|$.

Two independent sets of events are generated in a
pseudo-expe\-ri\-ment, one for the construction of templates and one to
define a pseudo-data sample.
These pseudo-data are analyzed, using the relevant templates, to estimate the model parameters. 
Repetition of the pseudo-experiments has demonstrated 
that the fit method is able to measure the model parameters in Eq.(\ref{fitpar})
within a wide range of initial values. The estimators obtained from
the fits are unbiased and have pull distribution dispersions close to one.

\section{Results}
\label{sec:Results}

\sloppy

\begin{figure*}[!hbt]
\begin{center}
\subfigure[]{
\includegraphics[width=0.35\textwidth]{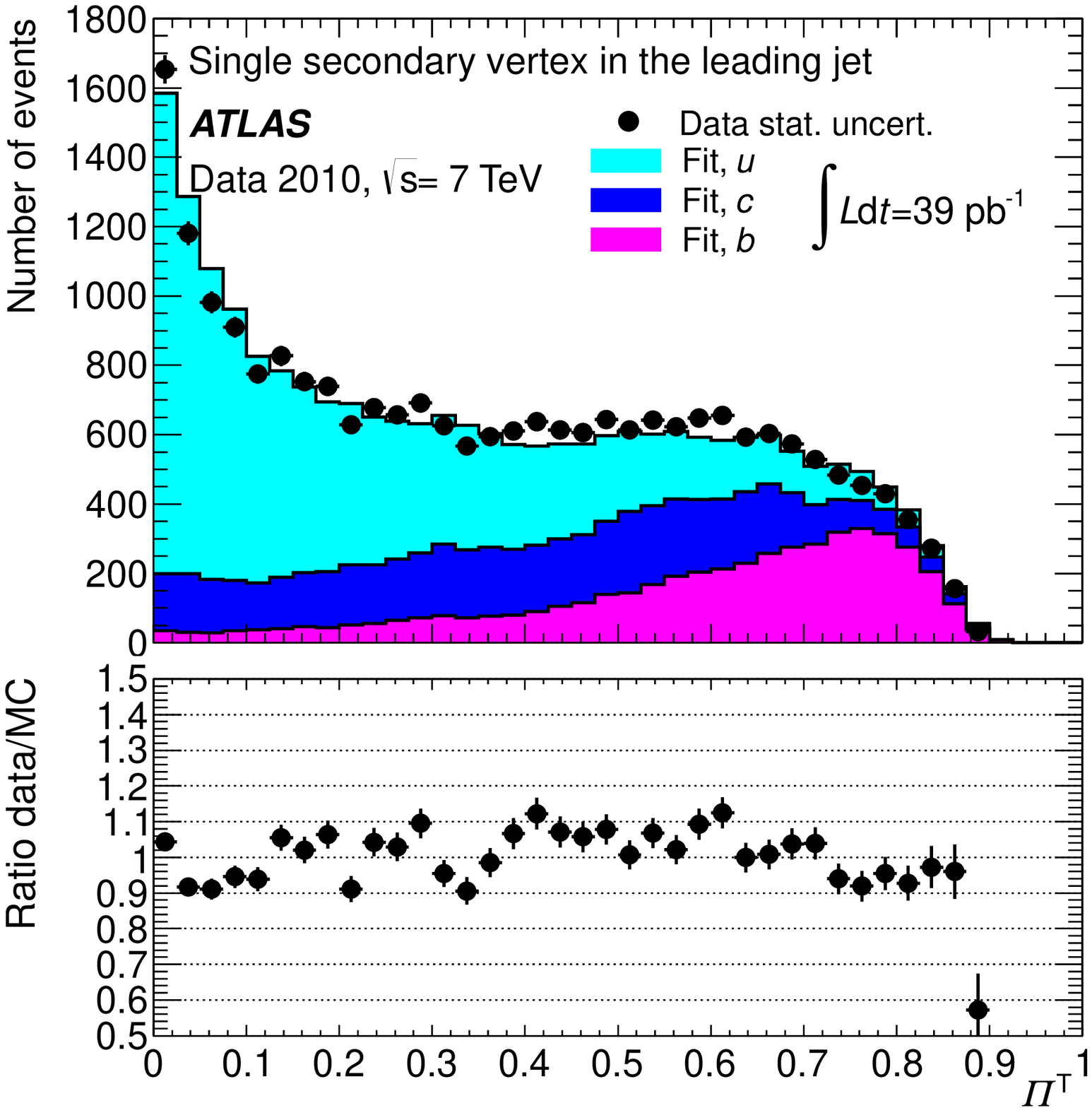}
}
\subfigure[]{
\includegraphics[width=0.35\textwidth]{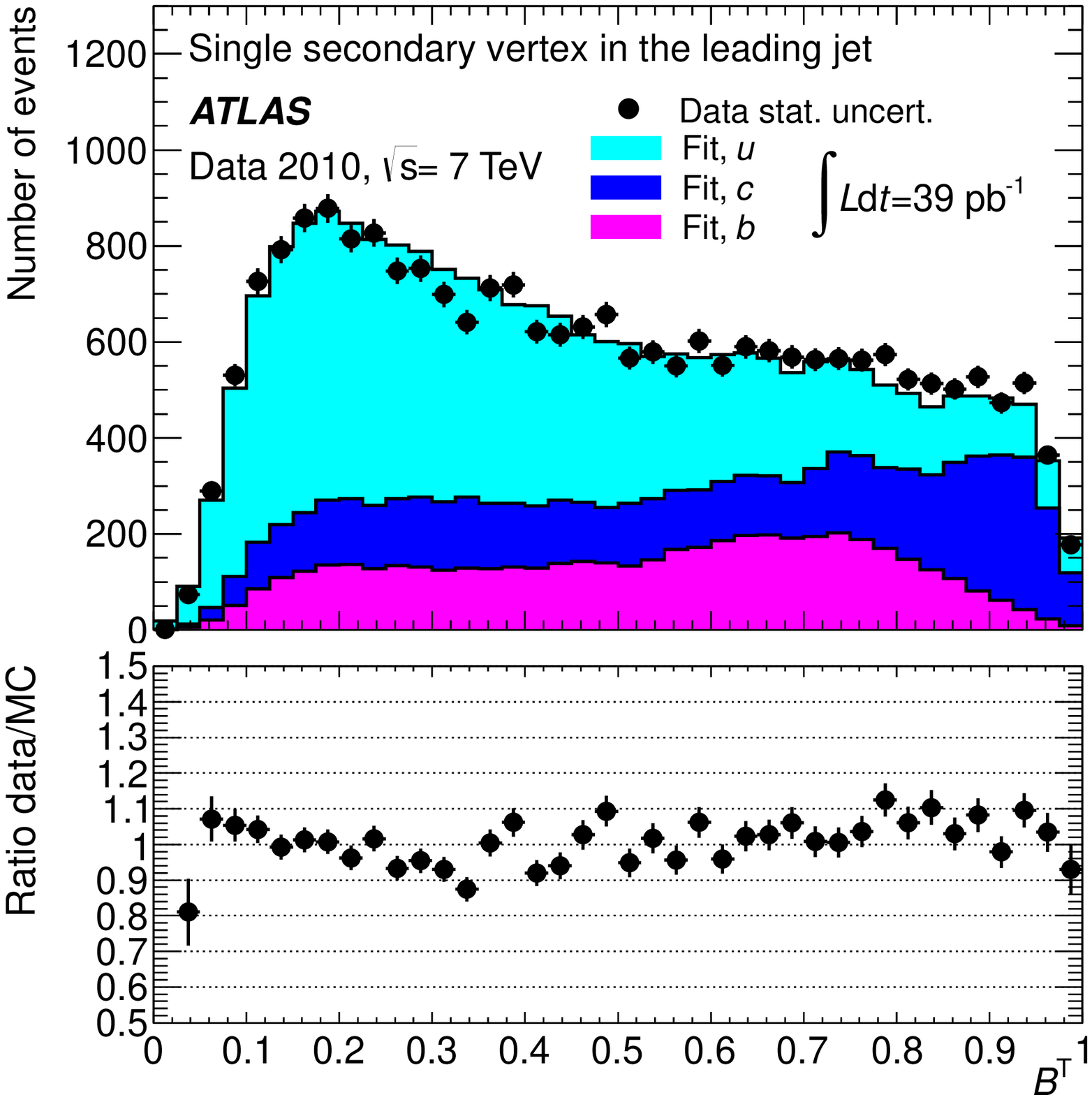}
}
\subfigure[]{
\includegraphics[width=0.35\textwidth]{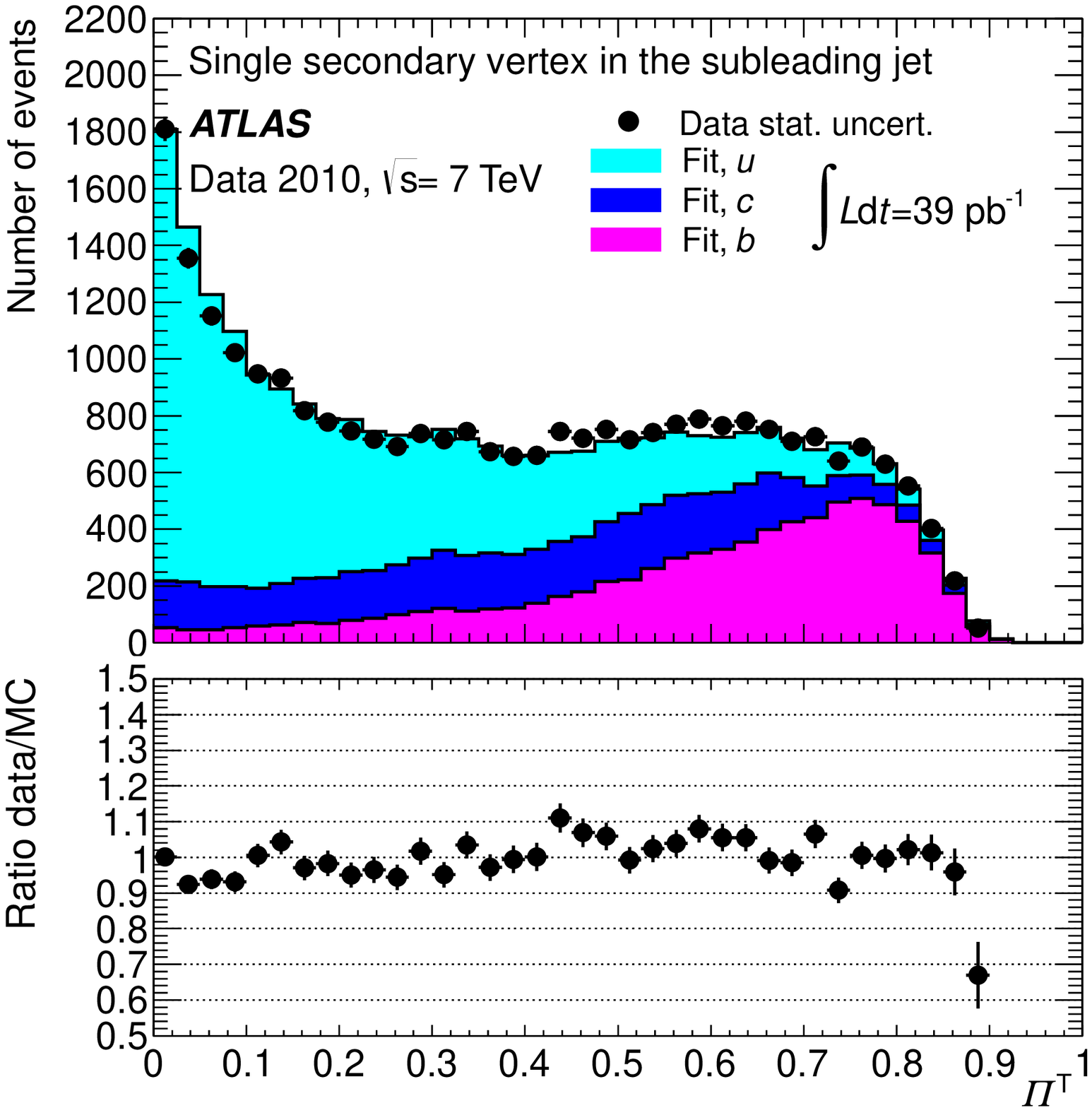}
}
\subfigure[]{
\includegraphics[width=0.35\textwidth]{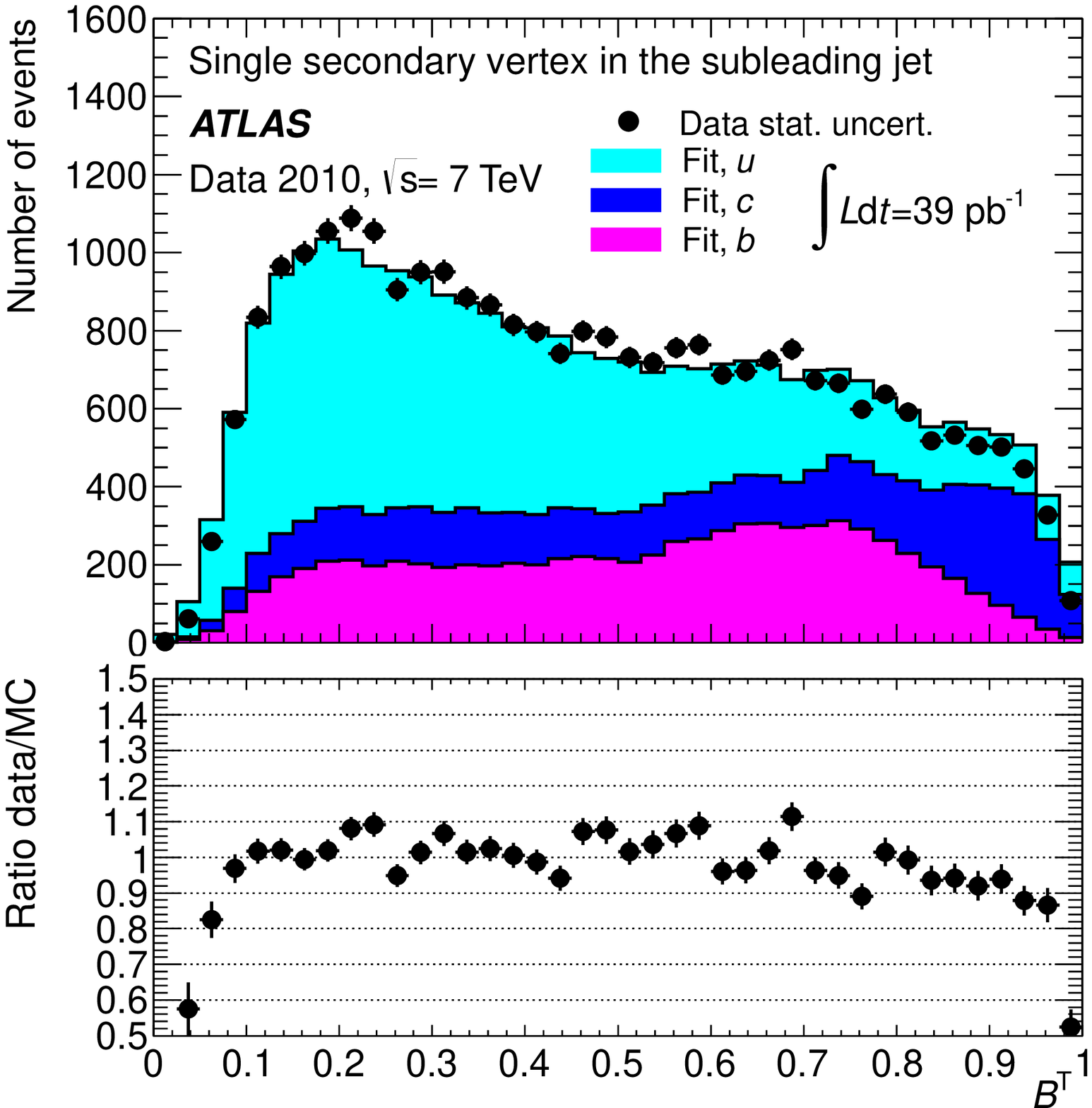}
}
\subfigure[]{
\includegraphics[width=0.35\textwidth]{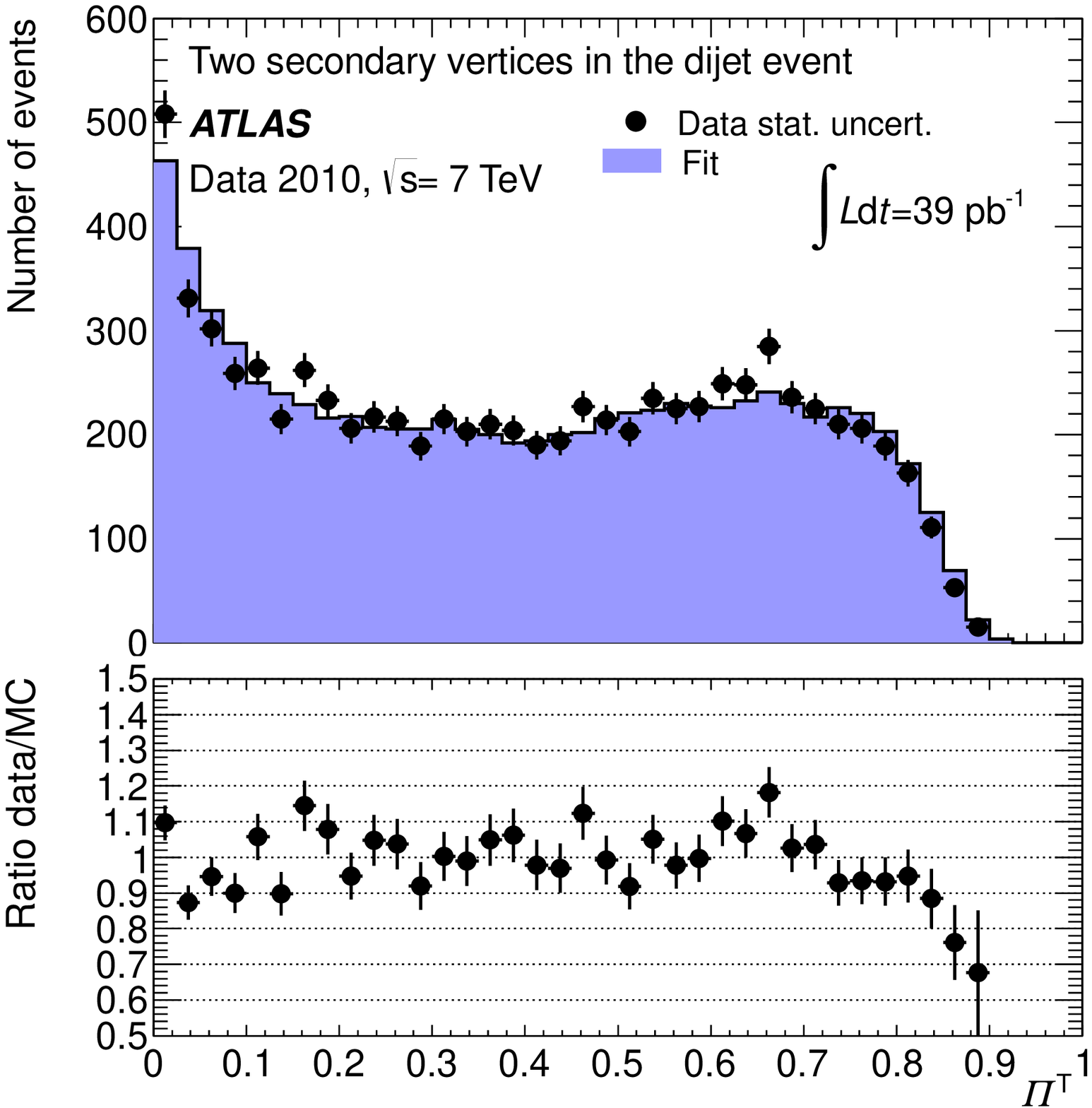}
}
\subfigure[]{
\includegraphics[width=0.35\textwidth]{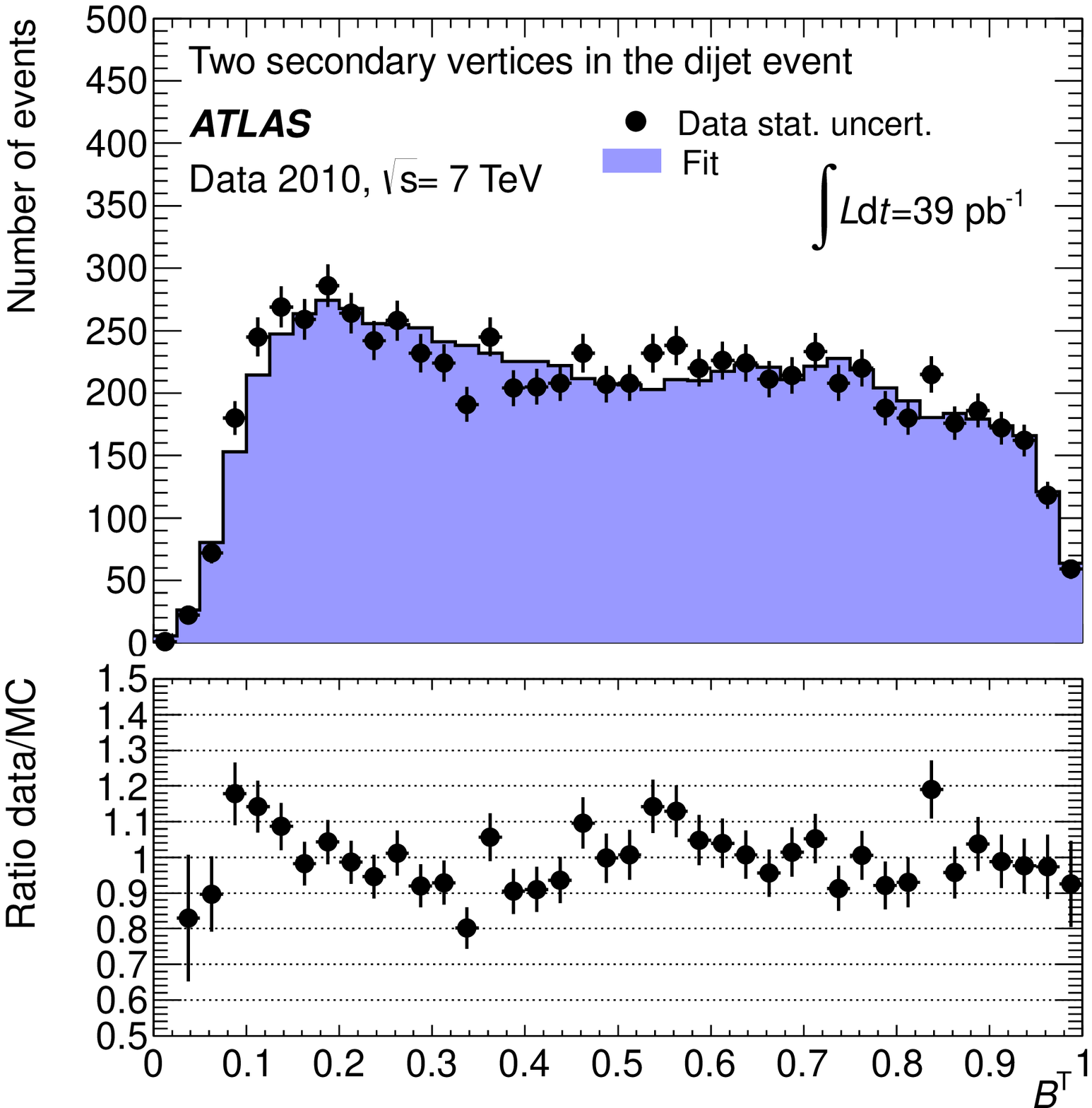}
}
  \caption{
     Data description with the Monte Carlo templates obtained as a result of the fit in the $[160,250]\GeV$ analysis bin.
    (a) and (b) show the $\Pi^{\top}$ and $B^{\top}$ distributions for secondary vertices in the leading jet, 
    (c) and (d) show the same distributions for secondary vertices in the subleading jet 
    in events with a single secondary vertex in two jets. (e) and (f) show the $\Pi^{\top}$ and $B^{\top}$ distributions in the events 
    with two secondary vertices averaged over leading and subleading jets.
    Data statistical uncertainties only are used to calculate the errors of the data to the Monte Carlo prediction ratios.
    } 
\label{figures:FitQuality}
\end{center}
\end{figure*}

\subsection{Data fit results}

An event-based extended maximum likelihood fit is used to fit the data. 
The fit is performed using the {\sc Minuit}~\cite{minuit} package included in the {\sc Root}~\cite{Brun199781} framework. 
A multi-nomial distribution is used in the likelihood function to describe the numbers of dijet events with zero, one or two reconstructed vertices. 
Using the {\sc Minuit} package, a detailed investigation of the
likelihood function in the region around its maximum value has been performed, to 
estimate the statistical uncertainties. It has been found that the parabolic approximation of the analysis fitting function is valid
around the maximum point.  

The quality of the description of the data obtained with the fit is illustrated 
in Fig.~\ref{figures:FitQuality}, where the data are compared with the Monte Carlo distributions predicted by the fit in the $[160,250]\GeV$ analysis bin.
All features of the data distribution are correctly reproduced, with a relative accuracy of better than 10\%.
The residual differences are within the systematic uncertainties of the measurements.

\begin{figure*}[!hbt]
\begin{center}
\subfigure[]{
\includegraphics[width=0.3\textwidth]{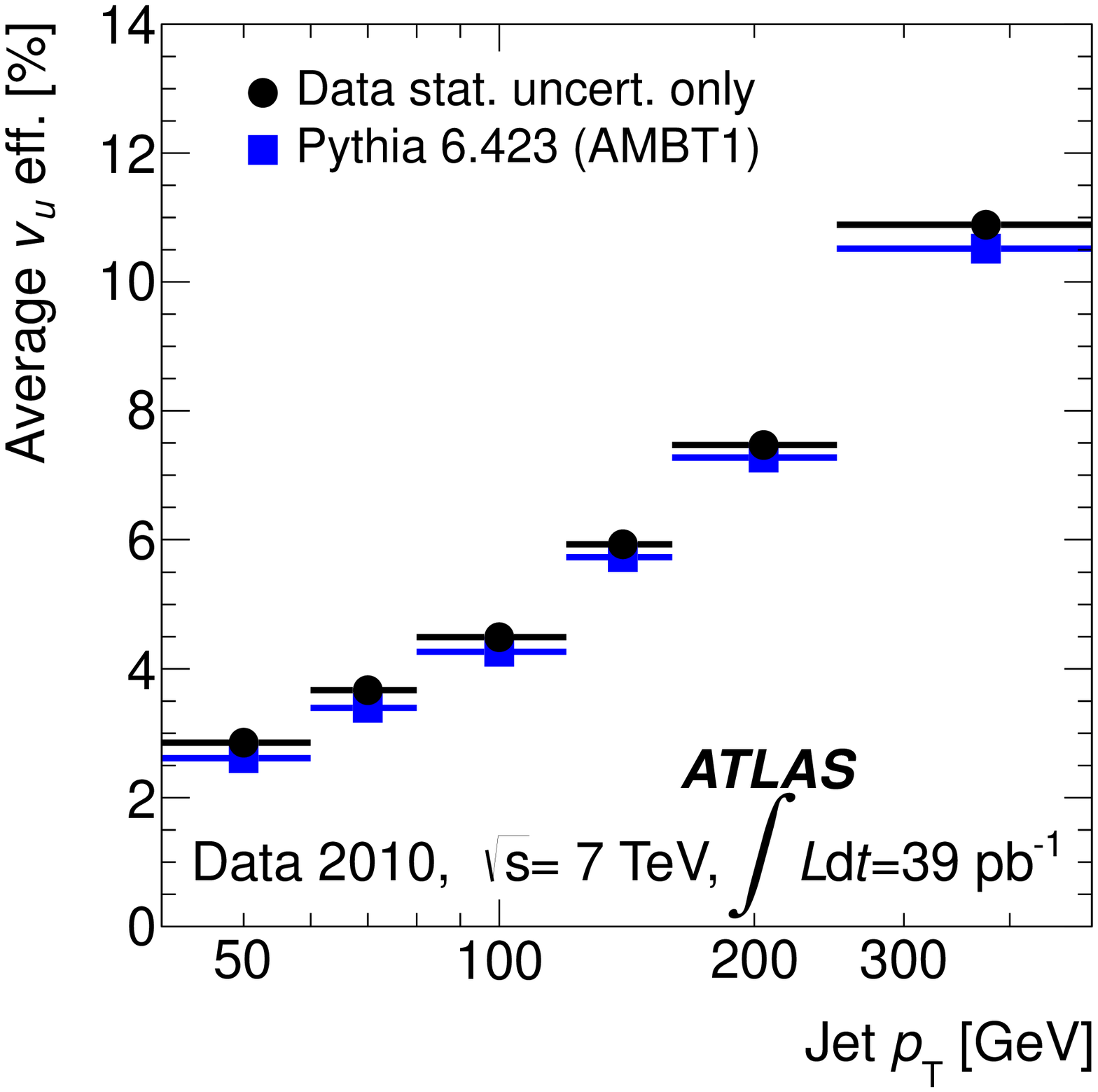}
}
\subfigure[]{
\includegraphics[width=0.3\textwidth]{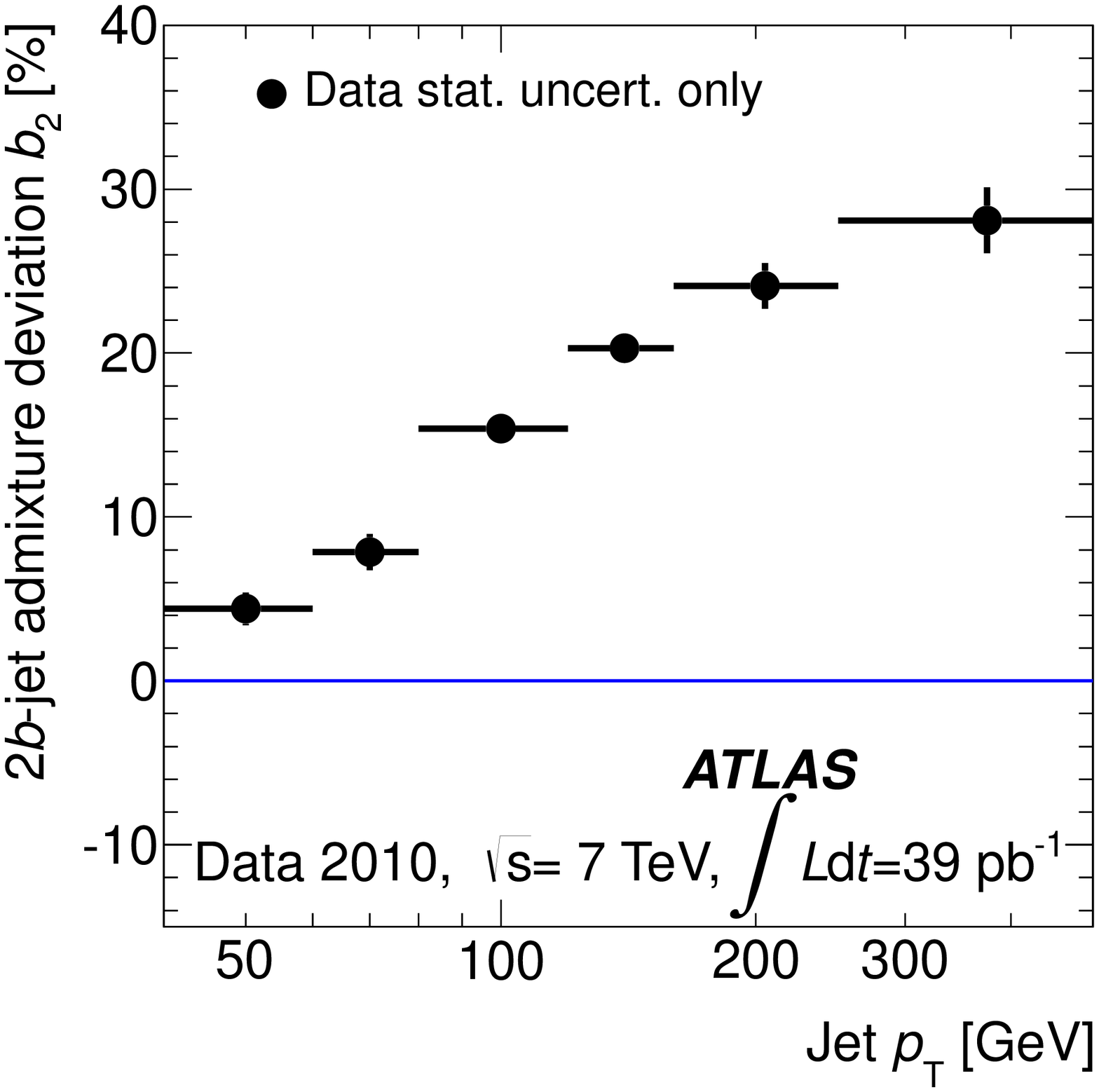}
}
\subfigure[]{
\includegraphics[width=0.3\textwidth]{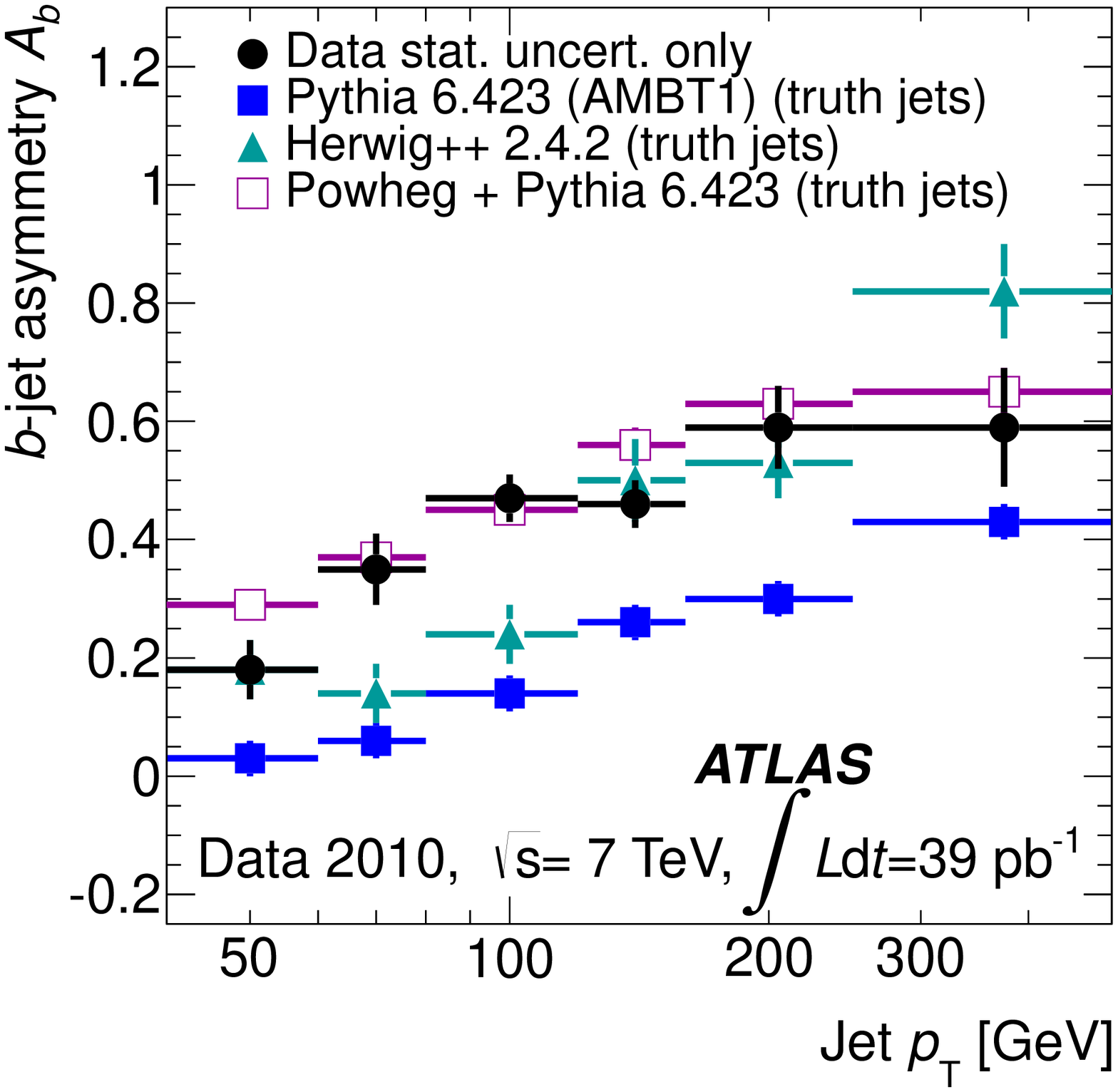}
}
  \caption{Data fit results for the (a) average fake vertex probability in light jets $v_u$ , 
      (b) $2b$-jet admixture deviation $b_2$ and (c) bottom dijet asymmetry $A_b$. 
   Statistical uncertainties only are shown.
   The fake vertex probability is shown with the {\sc Pythia}~6.423 reconstructed jet predictions.
   The $2b$-jet admixture deviation parameter should be zero if {\sc Pythia}~6.423 were fully consistent with data.
   The fitted bottom dijet asymmetry is corrected to the truth-particle jet level and compared with 
   {\sc Pythia}~6.423, Herwig++~2.4.2 and {\sc Powheg}+{\sc Pythia}~6.423 truth-particle jet predictions.  
 } 
\label{figures:FitResbis}
\end{center}
\end{figure*}

Figure~\ref{figures:FitResbis}(a) presents the fitted vertex probability in light jets together
with the prediction for dijet events generated with {\sc Pythia}~6.423 and passing through the full detector simulation.
The probability is averaged over leading and subleading jets in each $\pT$ bin. 
The vertices found in light jets are mainly fake ones (Sect. 6), therefore their probability is very sensitive 
to the details of the track and vertex reconstruction.
Good agreement between data and Monte Carlo simulation demonstrates
that the ATLAS detector performance is well understood in the Monte Carlo simulation. 

Figure~\ref{figures:FitResbis}(b) shows the deviation of the admixture of jets containing two bottom hadrons, $b_2$, from the {\sc Pythia}~6.423 prediction. The significance of the measured admixture excess confirms the importance of this additional contribution 
of double-bottom jets for a correct description of the data.  
This observation agrees with the results of \cite{BShapeCDF}.
The double-bottom jets are produced  by the gluon splitting mechanism (Sect. 5).
However, the analysis is unable to determine if a contribution from this mechanism to the fraction of jets 
with a single bottom hadron (see Fig.~\ref{figures:BProdb}) is also enhanced in data.

The fit results for the $b$-jet asymmetry $A_b$ need to be corrected for detector effects, in order to represent truth-particle jets.
The necessary correction is defined as a difference between truth-particle jet and reconstructed jet asymmetries, 
averaged over all $\pT$ bins using {\sc Pythia}~6.423, Herwig++~2.4.2 and {\sc Pythia}+{\sc EvtGen} dijet events.
The resulting correction of 0.08 $\pm$ 0.02 units is added to the fit results. 
The corrected $b$-jet asymmetry is
compared to the truth-particle $b$-jet asymmetries in {\sc Pythia}~6.423, {\sc Powheg}+{\sc Pythia}~6.423 
and Herwig++~2.4.2 in Fig.~\ref{figures:FitResbis}(c).
{\sc Pythia}~6.423 predicts a much smaller $b$-jet asymmetry than observed in the data.  
Since semileptonic decays are well described in {\sc Pythia}~6.423, the undetected energy due to neutrinos and muons from these decays 
cannot be the main contributor to the observed $b$-jet asymmetry. 
Modifications of the {\sc Pythia}~6.423 generator, such as different proton structure functions or 
different bottom parton fragmentation functions, are unable to improve substantially the agreement between the data and Monte Carlo simulation.
The $b$-jet asymmetry predicted by Herwig++~2.4.2 grows faster with $\pT$ than for the data. 
The best description of the data is provided by the {\sc Powheg}+{\sc Pythia}~6.423 generator, suggesting that NLO accuracy is needed 
to reproduce the $b$-jet asymmetry reliably.

\subsection{Unfolding}\label{sec:Unfolding}

To allow for a comparison with theoretical predictions 
and to remove detector resolution and acceptance effects, 
the flavour fractions for data must be unfolded to the truth-particle jet level as defined in Sect.~\ref{sec:Predictions}.
A simple bin-by-bin correction method is used. The expected inaccuracy introduced by the
unfolding procedure itself is small in comparison with the measurement uncertainties.
The unfolding correction factors for each flavour combination and leading jet $\pT$ bin 
are determined as ratios of the  reconstructed dijet events with required jet flavours 
to the corresponding truth-particle dijet events (Sect.~\ref{sec:Predictions}) in a given bin.
They are calculated using the fully simulated {\sc Pythia}~6.432 dijet sample 
and are typically in the $60\%$--$100\,\%$ range, mainly because of the $\pT$ cut on the reconstructed subleading jet. 
The corrections are different for dijet flavour fractions in the same $\pT$ bin due to
semileptonic decays of heavy flavour hadrons and different jet energy distributions for light and heavy flavour subleading jets.

The truth-particle dijet flavour fractions in each analysis bin are calculated using the following formula:
\begin{equation} \label{fracunfold}
 f_i^{\,\mathrm{unfold}} =  \frac{ f_i / \varepsilon_i} { \sum_k {(f_k / \varepsilon_k)}},
\end{equation}
where $f_i$ is a flavour fraction obtained in the fit and $\varepsilon_i$ is the corresponding unfolding correction factor. 
The $f_i^{\,\mathrm{unfold}}$ does not coincide with the $f_i$ because all correction factors $\varepsilon_i$ in a given analysis bin 
are different, as explained earlier.
Usually $\varepsilon_i$ is smaller than one; therefore the normalisation in Eq.(\ref{fracunfold}) is needed. 
The unfolded flavour fractions for truth-particle dijet events defined in Sect.~\ref{sec:Predictions}
are presented in Table~\ref{tables:UnfoldFrac}, as well as in Fig.~\ref{figures:Unfolded}, for the different leading jet $\pT$ bins.

\begin{table*}[!ht]
 \begin{center}
 \noindent\makebox[\textwidth]{
 \begin{tabular}{ccccccc}
 \hline
 Lead. jet $\pT$ [\GeV]  &   40--60  &   60--80  &   80--120 &   120--160  &   160--250  &   250--500 \\
 \hline
$f_{BB}$ [\%] & 0.65$\pm$0.04$\pm$0.12&  0.63$\pm$0.04$\pm$0.11&  0.58$\pm$0.02$\pm$0.11&  0.61$\pm$0.03$\pm$0.10&  0.58$\pm$0.05$\pm$0.07 &  0.39$\pm$0.08$\pm$0.06\\
$f_{BC}$ [\%] & 0.49$\pm$0.15$\pm$0.18&  0.31$\pm$0.13$\pm$0.18&  0.53$\pm$0.08$\pm$0.19&  0.52$\pm$0.09$\pm$0.22&  0.28$\pm$0.17$\pm$0.24 &  0.93$\pm$0.36$\pm$0.24\\
$f_{CC}$ [\%] & 1.08$\pm$0.30$\pm$0.31&  1.51$\pm$0.29$\pm$0.33&  1.03$\pm$0.11$\pm$0.28&  0.86$\pm$0.13$\pm$0.24&  1.68$\pm$0.30$\pm$0.44 &  0.70$\pm$0.47$\pm$0.50\\
$f_{BU}$ [\%] & 4.07$\pm$0.14$\pm$0.45&  4.78$\pm$0.14$\pm$0.46&  5.43$\pm$0.08$\pm$0.54&  6.02$\pm$0.09$\pm$0.52&  6.55$\pm$0.17$\pm$0.42 &  6.69$\pm$0.29$\pm$0.52\\
$f_{CU}$ [\%] & 10.6$\pm$0.5$\pm$1.7  &  10.3$\pm$0.5$\pm$1.3  &  11.3$\pm$0.25$\pm$1.5 &  10.9$\pm$0.24$\pm$1.8 &  11.0$\pm$0.5$\pm$2.0   &  12.4$\pm$0.8$\pm$2.8\\
$f_{UU}$ [\%] & 83.1$\pm$0.6$\pm$2.0  &  82.4$\pm$0.5$\pm$1.7  &  81.2$\pm$0.3$\pm$1.8  & 81.1$\pm$0.3$\pm$2.0  &  80.0$\pm$0.6$\pm$2.4   & 78.9$\pm$0.9$\pm$3.6\\
 \hline
\end{tabular}}
\end{center}
\caption{The unfolded dijet flavour compositions for each leading jet $\pt$ bin, with statistical uncertainties as first entries and the full systematic uncertainties as second entries.}
\label{tables:UnfoldFrac}
\end{table*}

\subsection{Systematic uncertainties}
\label{sec:ResultsSysUncert}

 The measured dijet flavour fractions are subject to systematic uncertainties, due to the assumptions made in selecting the model parameters in Eq.(\ref{fitpar}) 
and the following effects:
\begin{itemize}
\item Reconstructed jets in data and Monte Carlo simulation may have different
  kinematic properties due to trigger requirements, jet energy scale (JES) uncertainties, 
      cleaning cuts in the data selection procedure and event pile-up.
\item Differences between data and Monte Carlo simulation in the
  template shapes are possible, despite the tuning of the template
  shape to the track resolution, and the adjustment of the fit to
  increase the fraction of jets with two $b$-quarks.
\item The JES uncertainty and differences in energy between light and heavy flavour jets influence the unfolding correction factors.
      The template shapes are also affected by the remaining $\pT$ dependence of the $B^{\top}$ variable. 
\item Imperfect description of bottom and charm hadron decay properties in Monte Carlo generators. 
\end{itemize}

The influence of the differences in the jet $\pT$ and rapidity distributions between data and Monte Carlo simulation on the analysis results is estimated 
by using {\sc Pythia}~6.423 templates obtained with and without the $\pT$ and rapidity reweighting, respectively.   
The differences in the results are taken as systematic
uncertainties. Both make only minor contributions to the full systematic uncertainties.
The influence of pile-up is estimated by adding minimum bias 
events to the {\sc Pythia}~6.423 dijet events and repeating the analysis procedure. The effect is found to be negligible.

A potential bias due to the incorrect modelling of the JES is estimated by varying the jet energy response by its uncertainty~\cite{JetEnergyAtlas}. 
Detailed studies have shown 
that the JES uncertainty is smallest in the central calorimeter region
($\abseta<0.8$) for jets with $\pT>60\GeV$, with values of $\sim2.5\,$\%, 
and that it is well below the $5\,$\% level for the whole kinematic range of this analysis. 
Both jets in a jet pair are varied simultaneously. An additional $b$-jet energy uncertainty is taken into account, and also applied for charm jets. 
Templates obtained from {\sc Pythia}~6.423 events with modified jet energies are used for the data fit. 
Due to the dependence of the parameterisation of the charm and bottom
vertex reconstruction efficiencies on jet $\pT$, these values
are modified following the jet energy scaling. The systematic
uncertainty due to the JES is estimated to be half of the difference
between the fit results with positive and negative variation of the jet energy. 
The JES uncertainty is one of the major systematic uncertainties for all flavour fractions.
In particular, for $f_{BU}$ and $f_{CU}$ it varies from absolute
values of $0.2\,$\% and $1.1\,$\%  
in the lowest $\pT$ bin, to $0.1\,$\% and $0.8\,$\% in the highest $\pT$ bin.

The charm and bottom secondary vertex reconstruction efficiencies are fixed in the analysis to the predictions 
for {\sc Pythia}~6.423 dijet events, as explained in Sect.~\ref{sec:Method}.
To estimate possible deviations of these efficiencies, several Monte Carlo 
generators are used. The influences of a different proton structure function set ({\sc Pythia}+CTEQ~6.6),
a different parton fragmentation function ({\sc Pythia}+Peterson), a different showering model (Herwig++), 
different charm and bottom hadron decays description ({\sc Pythia}+{\sc EvtGen}) and additional track impact parameter smearing have been studied.
Herwig++ shows the largest deviations in the secondary vertex reconstruction efficiency for bottom from the {\sc Pythia}~6.423 Monte Carlo.
The absolute difference is $\sim 6\,$\% in the lowest $\pT$ region, but decreases to $\sim 2\,$\% in the highest $\pT$ region.
In the case of charm, {\sc Pythia}+{\sc EvtGen} predicts the largest absolute deviations of $\sim 2\,$\% from {\sc Pythia}~6.423. 
Since the largest uncertainty in the vertex reconstruction efficiency comes from the fragmentation model (Herwig++) 
for bottom and from the charm hadron decay description ({\sc EvtGen}) for charm,
the deviations in the charm and bottom vertex efficiencies are treated as independent for the systematic study.
The systematic uncertainties in the flavour fractions are estimated by varying the charm and bottom
vertex reconstruction efficiencies in the data fit by their maximal deviations.
The uncertainty due to the bottom vertex efficiency is
comparable with the JES uncertainty for the flavour fractions with bottom, and small otherwise.    
Similarly, the systematic uncertainty driven by the charm vertex efficiency is important for the fractions with charm.

The influence of imperfections in the Monte Carlo template shapes is estimated in two ways.
The baseline templates are constructed from Monte Carlo jets passing the dijet selection procedure. 
Alternatively, one can use jets without a dijet selection.
The templates obtained in this way are biased, due to different
kinematic properties of the jets and changes in the contributions of the different heavy flavour production mechanisms.
The number of contributing jets is also significantly larger, which
makes these templates virtually independent from the baseline ones.
To extract the systematic uncertainty, the data fit is redone with the inclusive jet templates. 
The statistical fluctuations due to the independent templates are
reduced by smoothing the differences in the fit results, using a linear function fit 
over the whole analysis $\pT$ range with weights $\sqrt{N_i}$, where $N_i$ is the number of selected data events in bin $i$. 
The smoothed differences in the flavour fractions between data fits are taken as systematic uncertainties.
In absolute values, they vary from $0.08\,$\% in the lowest $\pT$ bin to $0.2\,$\% in the highest $\pT$ bin for the $f_{CC}$ fractions and from $0.06\,$\% to $1.3\,$\% for the $f_{CU}$ fractions.
This systematic uncertainty is significantly smaller for the other flavour fractions.

Another check of the influence of the template shape is made 
by generating templates using Herwig++ instead of {\sc Pythia}~6.423.
The dedicated simulation model described in Sect.~\ref{sec:MethodFastMC} is exploited for this study.
The {\sc Pythia}~6.423 fully simulated vertices are used for template creation, but pseudo-data are created with Herwig++ vertices. 
Then the standard analysis procedure is applied. The averaged values based on 200 pseudo-experiments are compared with the initial fast simulation model parameters and
the differences are considered as systematic uncertainties. 
Overall, the systematic uncertainty due to the template shapes
constitutes a large contribution to the full systematic uncertainty
for all flavour fractions, and is
similar in size to those from JES and secondary vertex reconstruction efficiencies.

The predictions of the Monte Carlo simulation for the amount of heavy flavour in the leading and subleading jets differ significantly from one generator to another, 
as can be seen in Fig.~\ref{figures:asymflav}.
In the current analysis the charm production asymmetry is fixed to the {\sc Pythia}~6.423 prediction. 
To determine the systematic effect of an imprecise description of the charm asymmetry,
the data fit is redone with the charm asymmetry values given by {\sc Powheg}+{\sc Pythia}~6.423 as shown in Fig.~\ref{figures:asymflav}. 
This systematic uncertainty reaches $\sim 40\,\%$ of the total uncertainty for the $f_{CC}$ fraction and $\sim 20\,\%$ for the $f_{BC}$  fraction in the high $\pT$ region,
in all other cases it is below $\sim 10\,\%$.
The admixture of jets with two charm quarks inside is also fixed to the {\sc Pythia}~6.423 prediction in the analysis. 
To determine the systematic effect due to this, the double-charm admixture is varied by a fixed value, equal to 1/3 of the  measured double-bottom jet admixture. 
This choice is justified by a comparison of the bottom and charm asymmetries in Fig.~\ref{figures:asymflav}, which are governed by similar QCD effects.
This systematic uncertainty becomes important for the $f_{CU}$ and $f_{BU}$ fractions for large $\pT$. 
In absolute values, it is $1.2\,$\% for $f_{CU}$ and $0.35\,$\% for $f_{BU}$ in the $[250,500]\GeV$ bin.

To improve the agreement between data and Monte Carlo simulation, the flavour template shapes are tuned on the 2010 data as described in Sect.~\ref{subsec:TemplateTune}. 
The systematic uncertainties due to this procedure are estimated by
repeating  
the full analysis using the fully smeared ($F_{\mathrm{smear}}=1.0$, see Sect.~\ref{subsec:TemplateTune}) {\sc Pythia}~6.423 dijet sample for template construction
and definition of the vertex reconstruction efficiencies. 
This systematic uncertainty is $\sim50\,\%$ of the total systematic uncertainty for the $f_{BU}$ fraction in the high $\pT$ region and significantly smaller in other cases.

The unfolding procedure for obtaining the dijet flavour fractions at
the truth-particle level is based on estimations of the dijet
reconstruction efficiencies from Monte Carlo simulation.
Systematic uncertainties on these are estimated using the differences in the unfolded flavour fractions calculated with the unfolding coefficients 
predicted by {\sc Pythia}~6.423 and Herwig++~2.4.2. 
The flavour dijet reconstruction efficiencies are calculated for each analysis $\pT$ bin and therefore also depend on the JES modelling. 
The changes in the unfolded flavour fractions due to the shifted jet energies are considered as the JES-induced unfolding systematic uncertainties.
In both cases, the differences in the unfolded flavour fractions have significant statistical fluctuations 
due to the fact that the number of Monte Carlo events used for the reconstruction efficiency estimation is limited.
The differences for each flavour fraction are therefore smoothed in the same
way as the template shape systematic uncertainty. 
In the low $\pT$ bins the systematic uncertainties due to the unfolding are comparable in size 
to the uncertainties from JES and template shapes for $f_{CC}$, $f_{BU}$ and  $f_{CU}$.
In all other cases they are relatively small.

\begin{figure*}[!hbt]
\begin{center}
\subfigure[]{
\includegraphics[width=0.3\textwidth]{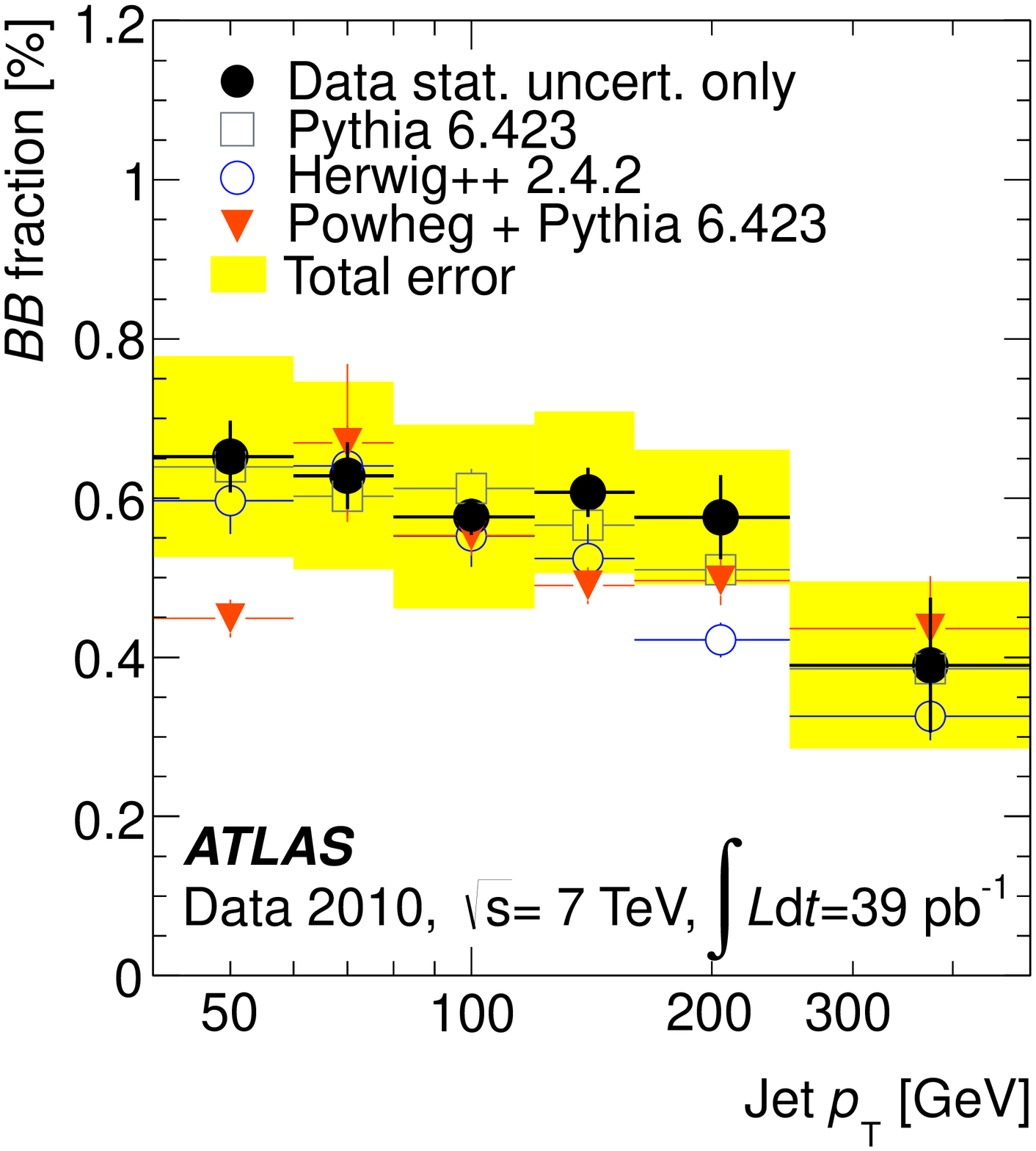}
\label{figures:Unfoldeda}
}
\subfigure[]{
\includegraphics[width=0.3\textwidth]{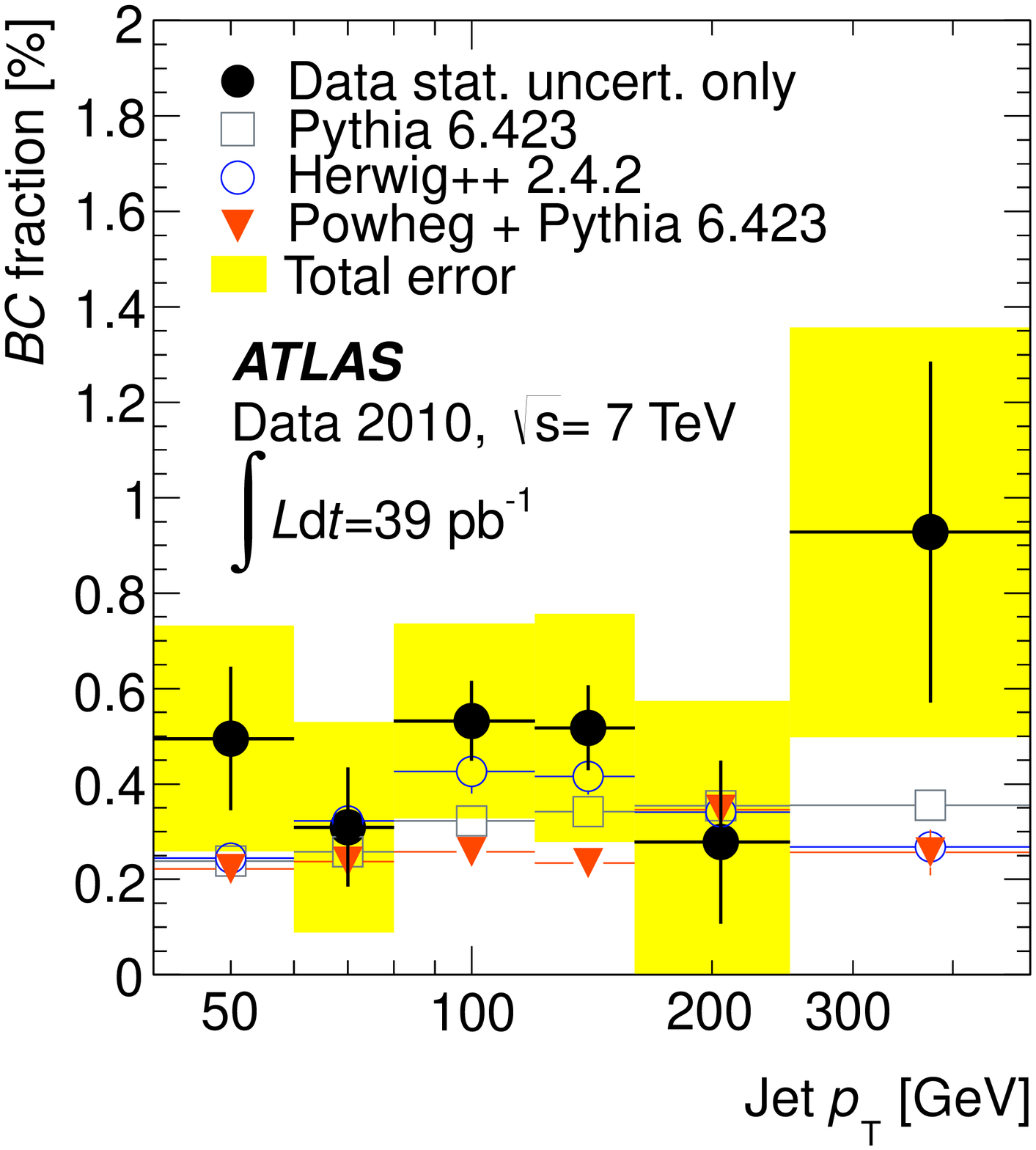}
\label{figures:Unfoldedb}
}
\subfigure[]{
\includegraphics[width=0.3\textwidth]{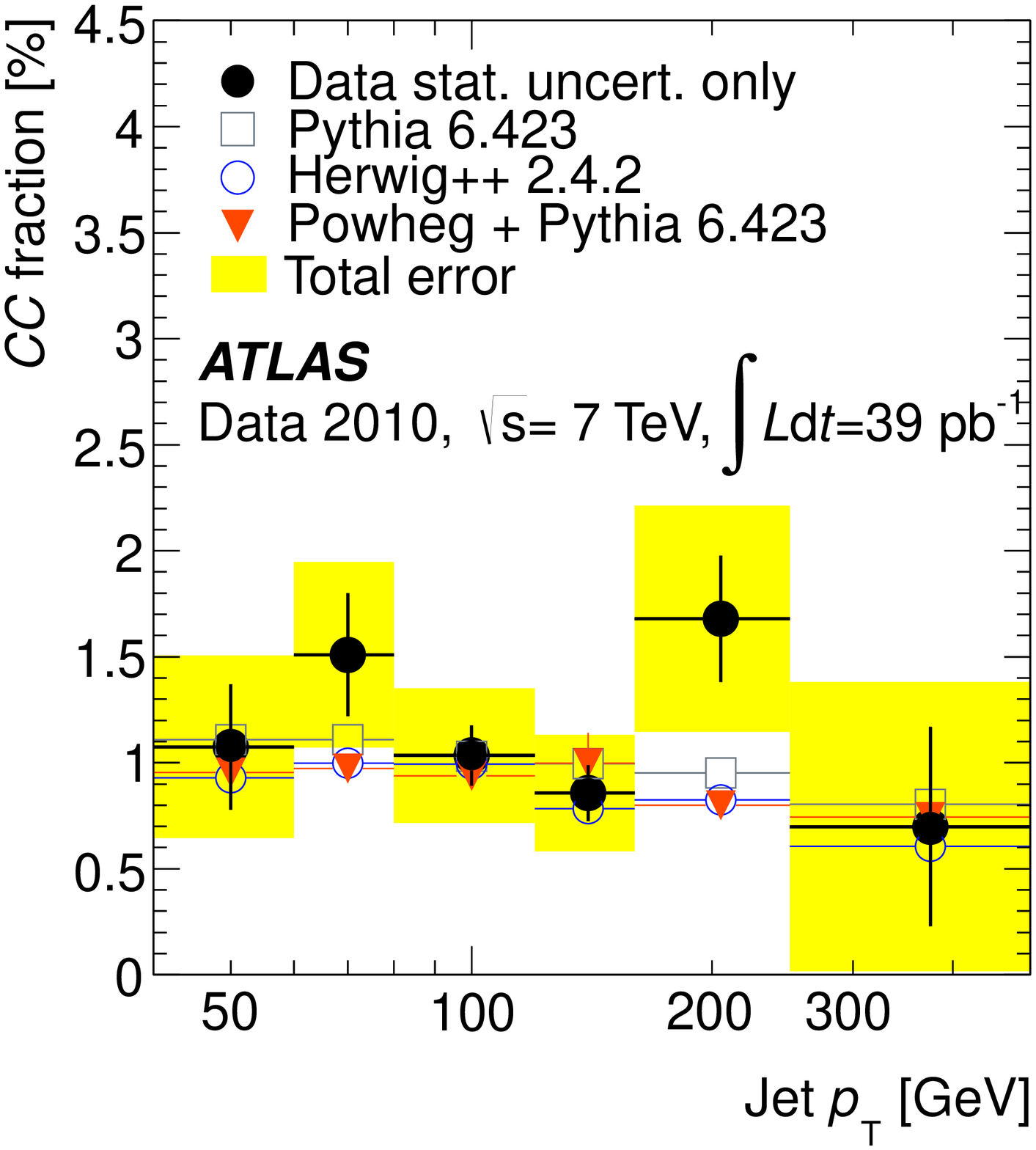}
\label{figures:Unfoldedc}
}
\subfigure[]{
\includegraphics[width=0.3\textwidth]{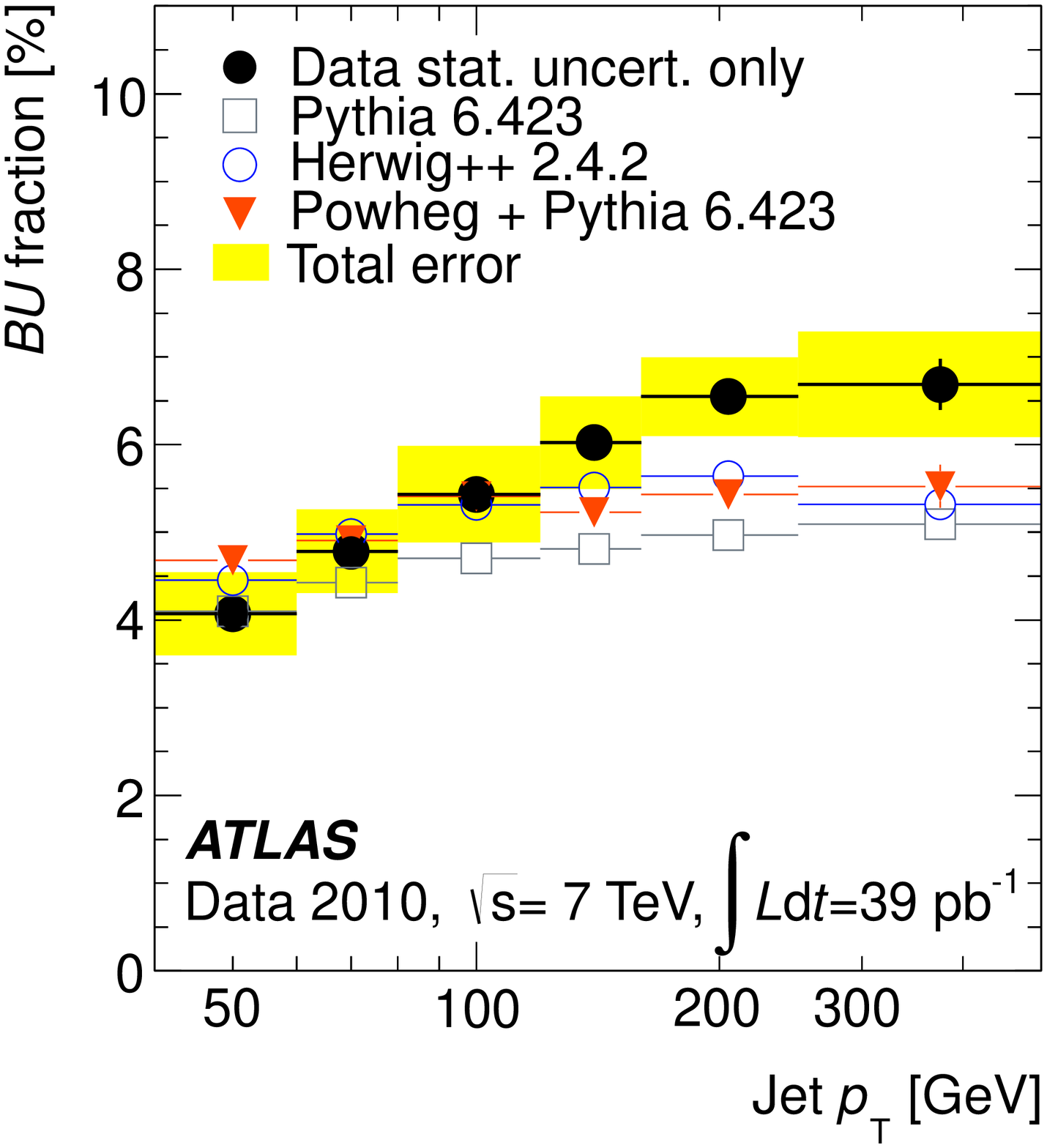}
\label{figures:Unfoldedd}
}
\subfigure[]{
\includegraphics[width=0.3\textwidth]{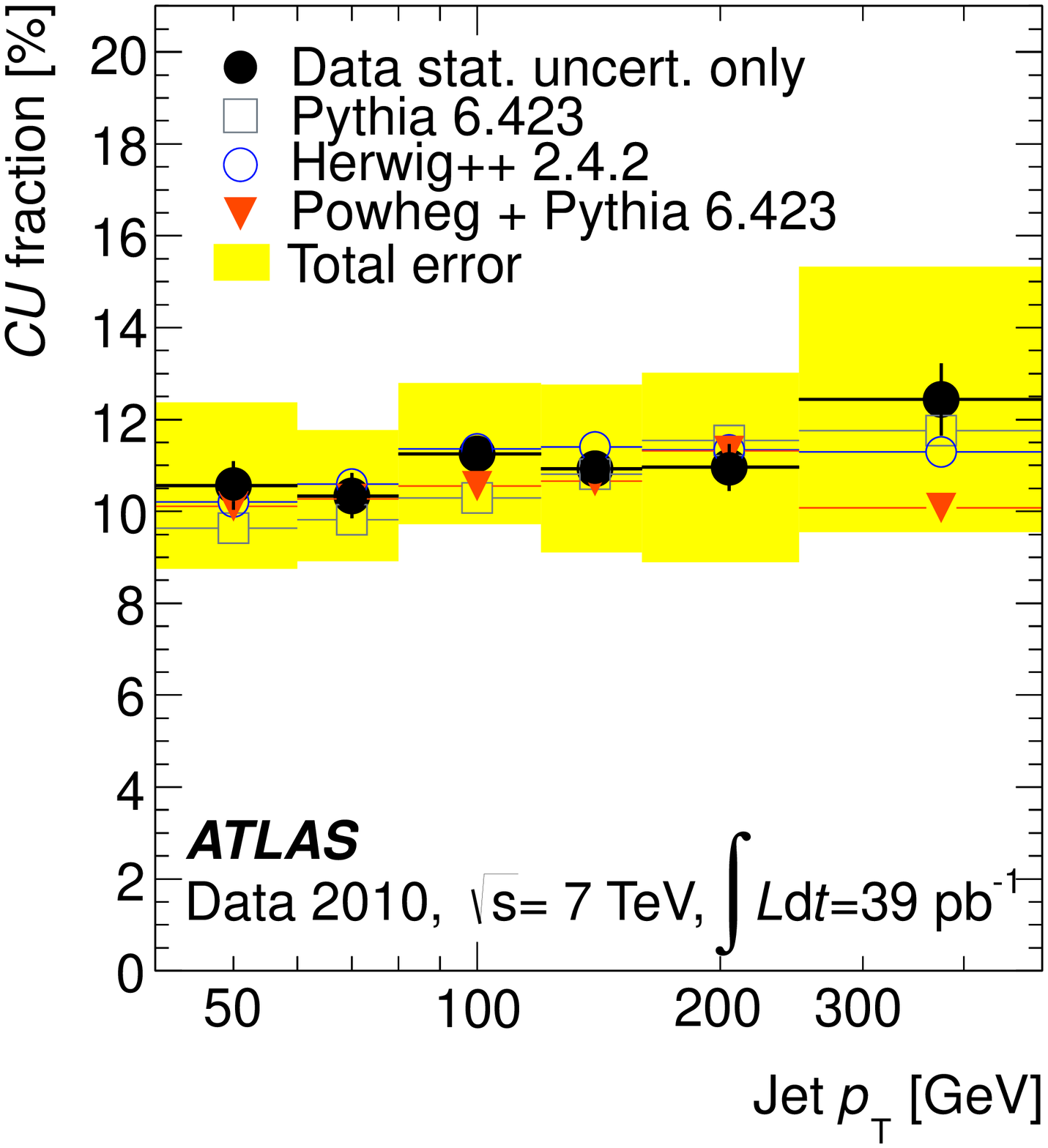}
\label{figures:Unfoldede}
}
\subfigure[]{
\includegraphics[width=0.3\textwidth]{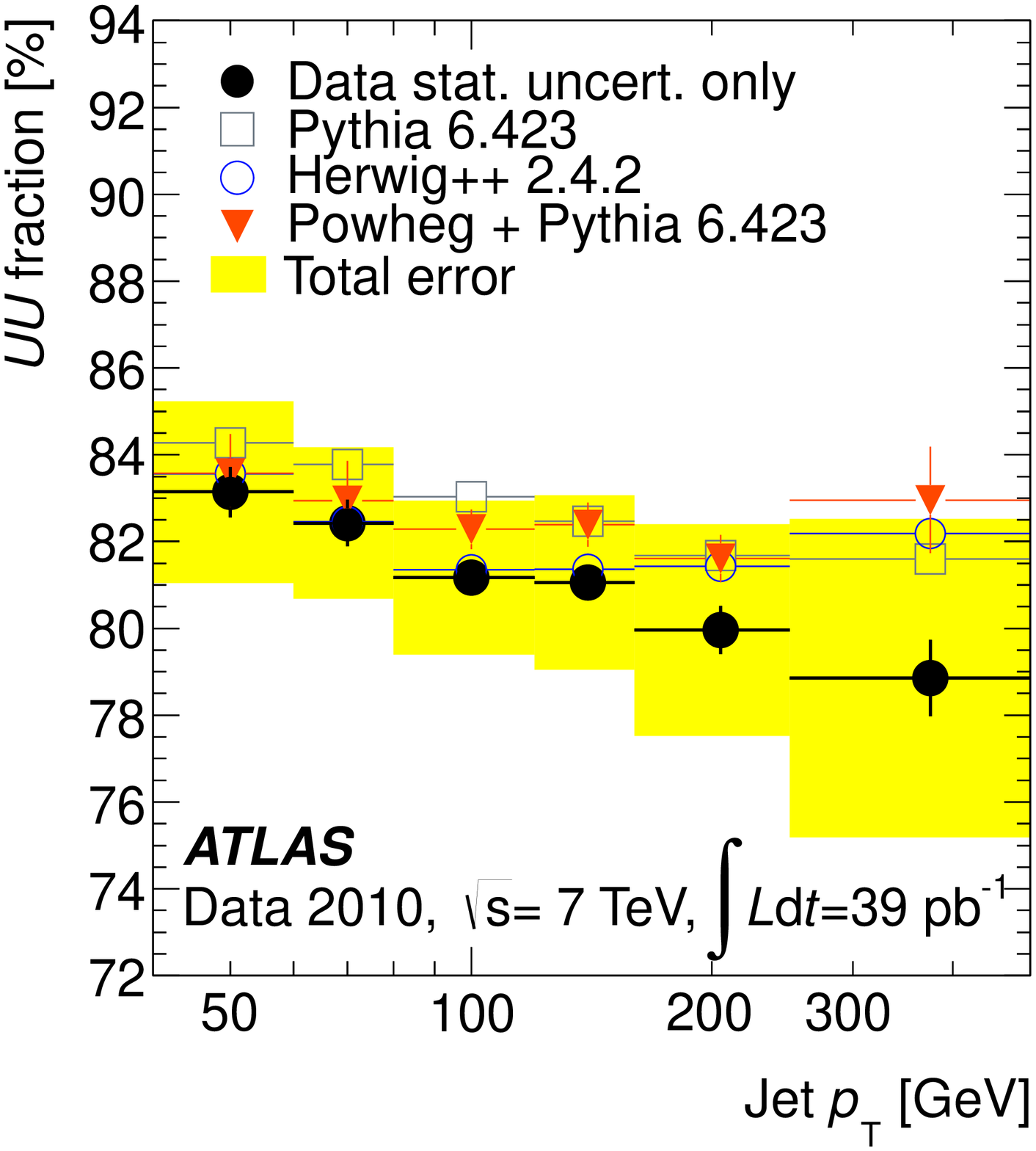}
\label{figures:Unfoldedf}
}
 \caption {The unfolded dijet flavour fractions for each leading jet $\pt$ bin (black points) with {\sc Pythia}~6.423 (squares), 
  Herwig++~2.4.2 (circles) and {\sc Powheg}+{\sc Pythia}~6.423 (filled triangles) 
  predictions overlaid. The error bars on the data points show statistical uncertainties only, whereas the full uncertainties appear as shaded bands.}
\label{figures:Unfolded}
\end{center}
\end{figure*}

The full systematic uncertainties on the unfolded dijet flavour fractions are presented in Table~\ref{tables:UnfoldFrac}.
These uncertainties are added in quadrature to the statistical uncertainties and are shown as shaded bands in Fig.~\ref{figures:Unfolded}. 
Except for $BU$, all data fractions are in agreement within the uncertainties with the predictions of the LO and NLO generators.
The $BU$ fraction, while coinciding reasonably well with the Monte Carlo simulation predictions at low jet $\pT$,
shows disagreement for jets with $\pT$ above $\sim100\GeV$. 
The discrepancy of the $BU$ data points with the {\sc Pythia}~6.423 prediction in the four high $\pT$ analysis bins 
has a significance of 4.3 standard deviations, corresponding to a fluctuation probability of $8.7 \times 10^{-6}$.

\fussy

\section{Conclusions}
\label{sec:Conclusion}
\sloppy
An analysis of the flavour composition of dijet events has been performed, based on an integrated luminosity of $39\,$~pb$^{-1}$ collected by the ATLAS detector 
in 2010 at a centre-of-mass energy of $7\TeV$. The analysis makes use of reconstructed secondary vertices in jets, without explicitly assigning 
individual flavours. Instead, kinematic properties of the ensemble of tracks associated with a secondary vertex are used to distinguish between light, charm and bottom jets. 
Specially constructed and optimised variables that are highly sensitive to the flavour content of jets, have been employed.
The dijet heavy flavour fractions are determined from a multidimensional fit using templates of these variables.

The analysis demonstrates the capability of ATLAS to measure the dijet fractions containing bottom jets
and the more challenging charm jets down to the level of $\sim0.5\,\%$. 
All five dijet final states with heavy flavours are reliably extracted and measured as a function of the leading jet $\pT$.

A significant difference in the bottom hadron content between leading and subleading jets is observed. 
This difference is poorly described by the LO generators {\sc Pythia}~6.423 and Herwig++~2.4.2, whereas the NLO generator {\sc Powheg} reproduces the data well.

The data-driven $b$-jet shape approach used in the fit demonstrates a deficiency of the $b$-jet template obtained 
with {\sc Pythia}~6.423, particularly in the high jet $\pT$ region. 
An increase of the template contribution describing the presence of
two $b$-hadrons inside a jet substantially improves the agreement
between data and Monte Carlo simulation. 

The measurements of the six dijet flavour fractions are compared with the predictions of the two LO 
generators {\sc Pythia}~6.423 and Herwig++~2.4.2, and also with the NLO generator {\sc Powheg}.  
All generator predictions are consistent with each other and agree
with the measured values, except for the mixed $BU$ dijet fraction, which is
systematically above all the predictions in the high $\pT$ region.

\fussy

\section{Acknowledgements}
\label{sec:Acknowledgements}
\sloppy
We thank CERN for the very successful operation of the LHC, as well as the
support staff from our institutions without whom ATLAS could not be
operated efficiently.

We acknowledge the support of ANPCyT, Argentina; YerPhI, Armenia; ARC,
Australia; BMWF, Austria; ANAS, Azerbaijan; SSTC, Belarus; CNPq and FAPESP,
Brazil; NSERC, NRC and CFI, Canada; CERN; CONICYT, Chile; CAS, MOST and NSFC,
China; COLCIENCIAS, Colombia; MSMT CR, MPO CR and VSC CR, Czech Republic;
DNRF, DNSRC and Lundbeck Foundation, Denmark; EPLANET and ERC, European Union;
IN2P3-CNRS, CEA-DSM/IRFU, France; GNAS, Georgia; BMBF, DFG, HGF, MPG and AvH
Foundation, Germany; GSRT, Greece; ISF, MINERVA, GIF, DIP and Benoziyo Center,
Israel; INFN, Italy; MEXT and JSPS, Japan; CNRST, Morocco; FOM and NWO,
Netherlands; RCN, Norway; MNiSW, Poland; GRICES and FCT, Portugal; MERYS
(MECTS), Romania; MES of Russia and ROSATOM, Russian Federation; JINR; MSTD,
Serbia; MSSR, Slovakia; ARRS and MVZT, Slovenia; DST/NRF, South Africa;
MICINN, Spain; SRC and Wallenberg Foundation, Sweden; SER, SNSF and Cantons of
Bern and Geneva, Switzerland; NSC, Taiwan; TAEK, Turkey; STFC, the Royal
Society and Leverhulme Trust, United Kingdom; DOE and NSF, United States of
America.

The crucial computing support from all WLCG partners is acknowledged
gratefully, in particular from CERN and the ATLAS Tier-1 facilities at
TRIUMF (Canada), NDGF (Denmark, Norway, Sweden), CC-IN2P3 (France),
KIT/GridKA (Germany), INFN-CNAF (Italy), NL-T1 (Netherlands), PIC (Spain),
ASGC (Taiwan), RAL (UK) and BNL (USA) and in the Tier-2 facilities
worldwide.

\fussy

\bibliographystyle{spphys}       
\bibliography{DijetFlavour}

\begin{thebibliography}{10}
\providecommand{\url}[1]{{#1}}
\providecommand{\urlprefix}{URL }
\expandafter\ifx\csname urlstyle\endcsname\relax
  \providecommand{\doi}[1]{DOI \discretionary{}{}{}#1}\else
  \providecommand{\doi}{DOI \discretionary{}{}{}\begingroup
  \urlstyle{rm}\Url}\fi

\bibitem{BBarD0}
{D0 Collaboration}, B.~Abbott, et~al., {The $b\bar{b}$ Production Cross Section
  and Angular Correlations in $p\bar{p}$ Collisions at $\sqrt{s}=1.8\TeV$},
  Phys. Lett. B \textbf{487}, 264 (2000)

\bibitem{BInclD0}
{D0 Collaboration}, B.~Abbott, et~al., {Cross Section for $b$ Jet Production in
  $p\bar{p}$ Collisions at $\sqrt{s}=1.8\TeV$}, Phys. Rev. Lett. \textbf{85},
  5068 (2000)

\bibitem{BBbarCDF}
{CDF Collaboration}, T.~Aaltonen, et~al., {Measurement of Correlated $b\bar{b}$
  Production in $p\bar{p}$ Collisions at $\sqrt{s}=1960\GeV$}, Phys. Rev. D
  \textbf{77}, 072004 (2008)

\bibitem{BBarCDFConf}
S.~Seidel, {Heavy Quark Production at the Tevatron}, arXiv:0808.3347 [hep-ex],
  (2008)

\bibitem{BBarSPS}
I.~Kenyon, {The Measurement of the Cross Section for Beauty Production at the
  CERN Anti-$p$ $p$ Collider}, J. Phys. G \textbf{15}, 1087 (1989)

\bibitem{BBarReview}
C.~Lourenco, H.~Wohri, {Heavy Flavour Hadroproduction from Fixed-Target to
  Collider Energies}, Phys. Rept. \textbf{433}, 127 (2006)

\bibitem{BInclCMS}
{CMS Collaboration}, {Inclusive $b$-Jet Production in $pp$ Collisions at
  $\sqrt{s}=7\TeV$}, J. High Energy Phys. \textbf{1204}, 084 (2012)

\bibitem{BBarCMS}
{CMS Collaboration}, {Measurement of the Cross Section for Production of
  $b\bar{b}X$ Decaying to Muons in $pp$ Collisions at $\sqrt{s}=7\TeV$}, J.
  High Energy Phys. \textbf{1206}, 110 (2012)

\bibitem{BHadCMS}
{CMS Collaboration}, {Inclusive $b$-Hadron Production Cross Section with Muons
  in $pp$ Collisions at $\sqrt{s}=7\TeV$}, J. High Energy Phys. \textbf{1103},
  090 (2011)

\bibitem{BBAngCMS}
{CMS Collaboration}, {Measurement of $B\bar{B}$ Angular Correlations Based on
  Secondary Vertex Reconstruction at $\sqrt{s}=7\TeV$}, J. High Energy Phys.
  \textbf{1103}, 136 (2011)

\bibitem{BBarLHCb}
{LHCb Collaboration}, {Measurement of $\sigma(pp \to b \bar{b} X)$ at
  $\sqrt{s}=7\TeV$ in the Forward Region}, Phys. Lett.B \textbf{694}, 209
  (2010)

\bibitem{BBarAtlasPaper}
{ATLAS Collaboration}, {Measurement of the Inclusive and Dijet Cross Sections
  of $b$-Jets in $pp$ Collisions at $\sqrt{s}=7\TeV$ with the ATLAS Detector},
  Eur. Phys. J. C \textbf{71}, 1846 (2011)

\bibitem{AtlasDetectorPaper}
{ATLAS Collaboration}, {The ATLAS experiment at the CERN Large Hadron
  Collider}, JINST \textbf{3}, S08003 (2008)

\bibitem{anti_kt}
M.~Cacciari, G.P. Salam, G.~Soyez, {The Anti-$k_t$ Jet Clustering Algorithm},
  J. High Energy Phys. \textbf{0804}, 063 (2008)

\bibitem{ATLAS-CONF-2010-038}
{ATLAS Collaboration}, {Data-Quality Requirements and Event Cleaning for Jets
  and Missing Transverse Energy Reconstruction with the ATLAS Detector in
  Proton-Proton Collisions at a Center-of-Mass Energy of $\sqrt{s}=7\TeV$},
  ATLAS-CONF-2010-038

\bibitem{JetEnergyAtlas}
{ATLAS Collaboration}, {Jet Energy Measurement with the ATLAS Detector in
  Proton-Proton Collisions at $\sqrt{s}=7\TeV$}, arXiv:1112.6426 [hep-ex],
  submitted to Eur. Phys. J. C

\bibitem{atlas_performance}
{ATLAS Collaboration}, {Expected Performance of the ATLAS Experiment -
  Detector, Trigger and Physics}, arXiv:0901.0512 [hep-ph],  (2009)

\bibitem{pythia64}
T.~Sjostrand, S.~Mrenna, P.Z. Skands, {{\sc Pythia}~6.4 Physics and Manual}, J.
  High Energy Phys. \textbf{0605}, 026 (2006)

\bibitem{Sherstnev:2007nd}
A.~Sherstnev, R.~Thorne, {Parton Distributions for LO Generators}, Eur. Phys.
  J. C \textbf{55}, 553 (2008)

\bibitem{ATLAS-CONF-2010-031}
{ATLAS}, {Charged Particle Multiplicities in $pp$ Interactions at
  $\sqrt{s}=0.9\TeV$ and $7\TeV$ in a Difractive Limited Phase-Space Measured
  with the ATLAS Detector at the LHC and New {\sc Pythia}~6 Tune},
  ATLAS-CONF-2010-031

\bibitem{herwigpp}
M.~Bahr, S.~Gieseke, M.~Gigg, D.~Grellscheid, K.~Hamilton, et~al.,
  {Herwig++~2.3 Release Note},   (2008)

\bibitem{Nadolsky:2008zw}
P.M. Nadolsky, H.L. Lai, Q.H. Cao, J.~Huston, J.~Pumplin, et~al., {Implications
  of CTEQ Global Analysis for Collider Observables}, Phys. Rev. D \textbf{78},
  013004 (2008)

\bibitem{herwig}
G.~Corcella, I.~Knowles, G.~Marchesini, S.~Moretti, K.~Odagiri, et~al.,
  {Herwig~6: An Event Generator for Hadron Emission Reactions with Interfering
  Gluons (Including Supersymmetric Processes)}, J. High Energy Phys.
  \textbf{0101}, 010 (2001)

\bibitem{jimmy1}
J.~Butterworth, J.R. Forshaw, M.~Seymour, {Multiparton Interactions in
  Photoproduction at HERA}, Z. Phys. C \textbf{72}, 637 (1996)

\bibitem{jimmy2}
J.~Butterworth, M.~Seymour.
\newblock {{\sc Jimmy}~4: Multiparton Interactions in Herwig for the LHC}.
\newblock {h}ttp://projects.hepforge.org/jimmy/ (2004)

\bibitem{ATLAS:1303025}
{ATLAS Collaboration}, First Tuning of Herwig/{\sc Jimmy} to ATLAS Data, Tech.
  Rep. ATL-PHYS-PUB-2010-014, CERN, Geneva (2010)

\bibitem{EvtGen}
D.~Lange, {The EvtGen Particle Decay Simulation Package}, Nucl. Instrum.
  Methods A \textbf{462}, 152 (2001)

\bibitem{pow}
S.~Alioli, P.~Nason, C.~Oleari, E.~Re, {A General Framework for Implementing
  NLO Calculations in Shower Monte Carlo Programs: the {\sc Powheg Box}}, J.
  High Energy Phys. \textbf{1006}, 043 (2010)

\bibitem{powbis}
P.~Nason, {A New Method for Combining NLO QCD with Shower Monte Carlo
  Algorithms}, J. High Energy Phys. \textbf{0411}, 040 (2004)

\bibitem{powter}
S.~Frixione, P.~Nason, C.~Oleari, {Matching NLO QCD Computations with Parton
  Shower Simulations: the {\sc Powheg} Method}, J. High Energy Phys.
  \textbf{0711}, 070 (2007)

\bibitem{powdij}
S.~Alioli, K.~Hamilton, P.~Nason, C.~Oleari, E.~Re, {Jet Pair Production in
  {\sc Powheg}}, J. High Energy Phys. \textbf{1104}, 081 (2011)

\bibitem{mstw2008}
A.~Martin, W.~Stirling, R.~Thorne, G.~Watt, {Parton Distributions for the LHC},
  Eur. Phys. J. C \textbf{63}, 189 (2009)

\bibitem{atlas_simulation}
{ATLAS Collaboration}, {The ATLAS Simulation Infrastructure}, Eur. Phys. J. C
  \textbf{70}, 823 (2010)

\bibitem{geant4}
{GEANT4}, S.~Agostinelli, et~al., {{\sc Geant}~4: A Simulation Toolkit}, Nucl.
  Instrum. Methods A \textbf{506}, 250 (2003)

\bibitem{Norrbin:2000zc}
E.~Norrbin, T.~Sjostrand, {Production and Hadronization of Heavy Quarks}, Eur.
  Phys. J. C \textbf{17}, 137 (2000)

\bibitem{Nakamura:2010zzi}
{Particle Data Group}, K.~Nakamura, et~al., {Review of Particle Physics}, J.
  Phys. G \textbf{37}, 075021 (2010)

\bibitem{BShapeCDF}
{CDF Collaboration}, T.~Aaltonen, et~al., {Measurement of $b$-Jet Shapes in
  Inclusive Jet Production in $p\bar{p}$ Collisions at $\sqrt{s}=1.96\TeV$},
  Phys. Rev. D \textbf{78}, 072005 (2008)

\bibitem{ATLAS-CONF-2010-070}
{ATLAS Collaboration}, {Tracking Studies for b-tagging with 7 TeV Collision
  Data with the ATLAS Detector}, ATLAS-CONF-2010-070

\bibitem{minuit}
F.~James, M.~Roos, {Minuit: A System for Function Minimization and Analysis of
  the Parameter Errors and Correlations}, Comput. Phys. Commun. \textbf{10},
  343 (1975)

\bibitem{Brun199781}
R.~Brun, F.~Rademakers, {\sc Root}: An Object Oriented Data Analysis Framework,
  Nucl. Instrum. Methods A \textbf{389}(1–2), 81 (1997)

\end{thebibliography}

\onecolumn 
\clearpage
\begin{flushleft}
{\Large The ATLAS Collaboration}

\bigskip

G.~Aad$^{\rm 48}$,
T.~Abajyan$^{\rm 21}$,
B.~Abbott$^{\rm 111}$,
J.~Abdallah$^{\rm 12}$,
S.~Abdel~Khalek$^{\rm 115}$,
A.A.~Abdelalim$^{\rm 49}$,
O.~Abdinov$^{\rm 11}$,
R.~Aben$^{\rm 105}$,
B.~Abi$^{\rm 112}$,
M.~Abolins$^{\rm 88}$
% ...
\bigskip
\\
% ...
$^{11}$ Institute of Physics, Azerbaijan Academy of Sciences, Baku, Azerbaijan\\
$^{12}$ Institut de F{\'\i}sica d'Altes Energies and Departament de F{\'\i}sica de la Universitat Aut{\`o}noma de Barcelona and ICREA, Barcelona, Spain\\
% ...
$^{21}$ Physikalisches Institut, University of Bonn, Bonn, Germany\\
% ...
$^{48}$ Fakult{\"a}t f{\"u}r Mathematik und Physik, Albert-Ludwigs-Universit{\"a}t, Freiburg, Germany\\
$^{49}$ Section de Physique, Universit{\'e} de Gen{\`e}ve, Geneva, Switzerland\\
% ...
$^{88}$ Department of Physics and Astronomy, Michigan State University, East Lansing MI, United States of America\\
% ...
$^{105}$ Nikhef National Institute for Subatomic Physics and University of Amsterdam, Amsterdam, Netherlands\\
% ...
$^{111}$ Homer L. Dodge Department of Physics and Astronomy, University of Oklahoma, Norman OK, United States of America\\
$^{112}$ Department of Physics, Oklahoma State University, Stillwater OK, United States of America\\
% ...
$^{115}$ LAL, Universit{\'e} Paris-Sud and CNRS/IN2P3, Orsay, France\\
% ...
\end{flushleft}

\end{document}